\documentclass[11pt, spanish]{article}
\usepackage{cite}
\usepackage{amsmath,amsfonts,amssymb}%,calrsfs}
\usepackage[small,bf,hang]{caption}
\usepackage{slashed}
\usepackage{color}

\usepackage{extarrows}
\usepackage{hyperref}

\usepackage{ulem}

\usepackage{geometry}
\def\hybrid{
        \topmargin -20pt
        \oddsidemargin 0pt
        \headheight 0pt \headsep 0pt
        \textwidth 6.25in % A4 paper
        \textheight 9.5in % A4 paper
        \marginparwidth .875in
        \parskip 5pt plus 1pt \jot = 1.5ex}

\hybrid

 \csname
@addtoreset\endcsname{equation}{section}

\usepackage{epsfig}
\usepackage{color}
\usepackage{graphics}
\usepackage{graphicx}
\usepackage{slashed}

\usepackage{amsthm}
\usepackage{amssymb}
\usepackage{amsmath}
\usepackage{amsfonts}

\numberwithin{equation}{section}

\newcommand{\be}{\begin{equation}}
\newcommand{\ee}{\end{equation}}
\newcommand{\ba}{\begin{eqnarray}}
\newcommand{\ea}{\end{eqnarray}}

\newcommand{\nn}{\nonumber}

\input xy
\xyoption{all}

\begin{document}
\begin{titlepage}
\rightline{}
%\rightline\today
\begin{center}
\vskip 1.5cm
 {\Large \bf{   Exploring the
 $\beta$ symmetry of supergravity}}
\vskip 1.7cm

{\large\bf {Walter H. Baron$^1$, Diego Marqu\'es$^{2,3} $ and Carmen A. N\'u\~nez$^2$}}
\vskip 1cm

$^1$ {\it  Instituto de F\'isica de La Plata}, (CONICET-UNLP)\\
{\it Departamento de Matem\'atica, Universidad Nacional de La Plata. }\\
{\it C. 115 s/n, B1900, La Plata, Argentina}\\
wbaron@fisica.unlp.edu.ar\\
 
\vskip .3cm

$^2$ {\it   Instituto de Astronom\'ia y F\'isica del Espacio}, (CONICET-UBA)\\
{\it Ciudad Universitaria, Pabell\'on IAFE, CABA, C1428ZAA, Argentina}\\
diegomarques@iafe.uba.ar, \ \ carmen@iafe.uba.ar\\

\vskip .3cm

$^3$ {\it Departamento de F\'isica, Universidad de Buenos Aires.}\\
{\it Ciudad Universitaria, Pabell\'on 1, CABA, C1428ZAA, Argentina}\\

\vskip .1cm

\vskip .4cm

\vskip .4cm

\end{center}
\bigskip\bigskip

\begin{center} 
\textbf{Abstract}

\end{center} 
\begin{quote}  
Kaluza-Klein reductions of low energy string effective actions possess a continuous $O(d,d) $ symmetry. The non-geometric elements of this group, parameterized by a bi-vector $\beta$, are not inherited from the symmetries of the higher-dimensional theory, but constitute instead a symmetry enhancement produced by the isometries of the background. The realization of this enhancement in the parent  theory was recently defined as $\beta$ symmetry, a powerful tool that allows to avoid the field reparameterizations of the Kaluza-Klein procedure. In this paper we further  explore this symmetry and its impact on the first order $\alpha'$-corrections. We derive the $\beta$ transformation rules from the frame formulation of Double Field Theory (DFT), and connect them to the corresponding rules in the  Metsaev-Tseytlin and Bergshoeff-de Roo supergravity schemes. 
It follows from our results that $\beta$ symmetry is a necessary condition for the uplift of string  $\alpha'$-expansions to DFT.

\end{quote} 
\vfill
\setcounter{footnote}{0}
\end{titlepage}

\tableofcontents

\section{Introduction}

The low energy limits of $D$-dimensional string theories  compactified on $T^d$  are invariant with respect to the global
$O(d,d; R)$ group \cite{sen}. There are two distinct types of elements in this group. On the one hand, the geometric subgroup $GL(d) \times R^{\frac{d(d-1)}2}$ descends from the diffeomorphisms and two-form shifts of the higher dimensional theory. On the other hand, the complement of this subgroup is not associated to a symmetry of the higher dimensional theory, and so it constitutes a symmetry enhancement produced by the toroidal truncation. 

The compactification from the higher to the lower dimensional action is achieved through  Kaluza-Klein (KK) reduction.  When only the massless modes are kept, KK reductions involve two steps. First, the isometries of the background are imposed on the fields, which simply amounts to removing their dependence on the toroidal coordinates.
The second step is a rearrangement of degrees of freedom, namely field redefinitions. The fields of the parent action, which constitute the multiplets of the higher dimensional  diffeomorphisms, are decomposed into multiplets of the lower dimensional diffeomorphisms and gauge symmetries. While it is in the first step that the symmetry is enhanced to $O(d,d; R)$, the second step is intended to make this symmetry (and others) manifest. 

Recently, it was shown in \cite{bmn} that the non-geometric elements of $O(d,d; R)$, which are parameterized by a rigid bi-vector $\beta$, act covariantly on the multiplets of the higher dimensional diffeomorphisms. In other words, the $\beta$ transformations preserve the structure of the higher-dimensional fields, which permits to avoid the second step of the KK procedure. This in turn allows to act with $\beta$ transformations on the higher dimensional action, and to assess/verify its invariance under the assumption of isometries. Even if one is actually simply dealing with the $O(d,d;R)$ invariance of the lower dimensional theory, effectively this looks like a symmetry of the higher-dimensional action,  and so we dubbed it $\beta$ symmetry of supergravity. For all practical purposes, this can be conceived as a symmetry of the higher dimensional theory, whose effect in lower dimensions serves to enhance $GL(d) \times R^{\frac{d(d-1)}2}$ to the full $O(d,d;R)$ group. Interestingly, since this symmetry mixes the NSNS fields, it fixes their relative couplings and might then become a powerful tool to compute string theory interactions.

The $\beta$-invariance of the string effective actions to first order in $\alpha'$ was proved in \cite{bmn}. We presented the  $\beta$ transformation rules, their first order corrections, the closure of the symmetry algebra and the invariance of the Lagrangian,  in the so called generalized Bergshoeff-de Roo scheme \cite{marn}. In this paper, we further explore several features of the $\beta$ symmetry of string supergravities.

In {\bf Section 2} we derive $\beta$ symmetry from Double Field Theory (DFT) \cite{dft}. We use the background independent frame-like formulation of DFT  \cite{hk,Exploring}, restricted by the so-called strong constraint. The formalism is manifestly invariant under the global $O(D,D;R)$ duality group, which contains the $\beta$ elements already embedded into $GL(D)$, ready to act on the $D$-dimensional fields that parameterize the duality multiplets. Of course, they are precisely the elements that determine the $\beta$ symmetry of supergravity. Here we derive the $\beta$ transformation rules to first order in $\alpha'$ in the DFT scheme of supergravity,  following the standard procedure of solving the strong constraint by annihilating the dual derivatives, and gauge-fixing the double-Lorentz symmetry. In other words, since it is implied by DFT, $\beta$ symmetry is a necessary condition for the uplift of supergravity to DFT, at least to this formulation. As such, it can serve as a tool to understand potential obstructions to manifest duality covariant uplifts of supergravity interactions.

Before moving from the DFT scheme of supergravity to other more conventional schemes, we discuss various scheme-independent results in {\bf Section 3}. $\beta$ transformations to lowest order are unique, and preserve the Lagrangian exactly. The two-parameter first order corrections derived from DFT also leave the Lagrangian invariant and define a symmetry algebra that closes off-shell. We also display a useful four-parameter trivial  deformation of the first order transformations that leave the action invariant (i.e. the variation of the Lagrangian leads to a total derivative). While the symmetries that leave the Lagrangian invariant close off-shell, we show that the transformations that preserve the action but not the Lagrangian also close, but on-shell. These, and other scheme independent ambiguities in the structure of the first order corrections to $\beta$ symmetry are discussed in detail in this section.

While {\bf Section 4} is devoted to recover all the results of \cite{bmn}, by performing the field redefinitions that connect the DFT scheme with the generalized Bergshoeff-de Roo scheme, in {\bf Section 5} we show how $\beta$ symmetry works in the Metsaev-Tseytlin scheme. In all sections the results are presented for a bi-parametric theory, such that the bosonic and heterotic strings are obtained for specific choices of the parameters. In particular, we show that $\beta$ symmetry in the Metsaev-Tseytlin scheme  of the bosonic string admits a purely  metric formulation, as expected. Instead, the $\beta$-invariance of the Chern-Simons form in the heterotic string is necessarily frame-like. We discuss both cases separately, providing the transformation rules that leave the Lagrangian invariant, as well as the deformations that preserve the action, and the brackets of the symmetry algebra.

We end with some conclusions in {\bf Section 6}. Notation, definitions and side computations are contained in three Appendices.

\section{$\beta$ transformations from Double Field Theory}\label{dft}

In this section we derive the $\beta$ transformations of supergravity to first order in $\alpha'$ from their action in the strong constrained and background independent frame formulation of Double Field Theory (DFT) \cite{hk, Exploring}. The procedure requires solving the strong constraint in the supergravity section, and applying a gauge fixing of the double Lorentz symmetry, after which it systematically provides the $\beta$ transformation rules of the supergravity fields in the DFT scheme. Other supergravity schemes can then be reached from this one through field redefinitions. 

The field content of DFT involves  a pair of generalized frames $E_M{}^{\underline a}$ and $E_M{}^{\overline a}$ and a generalized dilaton $d$. The symmetries that we will deal with in this paper are $O(D,D)$, with invariant metric $\eta_{M N}$, and two independent Lorentz symmetries with invariant metrics $\eta_{\underline {a b}}$ and $\eta_{\overline {a b}}$, each one acting on its respective frame. All indices are raised and lowered with their corresponding $\eta$ invariant metric. The generalized frames are constrained to satisfy the identities
\be
E_{M \underline a} E^M{}_{\underline b} = \eta_{\underline {a b}} \ , \ \ \ E_{M \underline a} E^M{}_{\overline b} = 0 \ , \ \ \ E_{M \overline a} E^M{}_{\overline b} = \eta_{\overline {a b}}  \, .
\ee
The fields are formally defined on a doubled space and so are acted on by derivatives $\partial_M$, which are  strong constrained, meaning that the $O(D,D)$ invariant contraction of derivatives vanishes.

Infinitesimally, $O(D,D)$ acts through a constant antisymmetric matrix $h_{M N}$, and each Lorentz group through independent local antisymmetric parameters $\Lambda_{\underline {a b}}$ and $ \Lambda_{\overline {a b}}$. While the generalized dilaton is invariant, the two generalized frames transform as
\begin{subequations}
\begin{align}
\delta E_M{}^{\underline a} \ &= \ h_M{}^N E_N{}^{\underline a} + E_M{}^{\underline b} \Lambda_{\underline b}{}^{\underline a} + E_M{}^{\overline b} \Delta_{\overline b}{}^{\underline a}\, , \\
\delta E_M{}^{\overline a} \ &= \ h_M{}^N E_N{}^{\overline a} + E_M{}^{\overline b} \Lambda_{\overline b}{}^{\overline a} - E_M{}^{\underline b} \Delta^{\overline a}{}_{\underline b} \ .
\end{align}
\end{subequations}
The last terms in these expressions account for the generalized Green-Schwarz transformation \cite{marn,odd,dw,dw2,BaronDCFR}, consisting on all-order higher-derivative deformations of the double Lorentz transformations. In particular, the first order deformation is given by
\be
\Delta^{(1)}_{\overline a \underline b} = \frac a 2 F_{\overline a \underline{cd }} D_{\underline b} \Lambda^{\underline {c d}} + \frac b 2 D_{\overline a} \Lambda^{\overline {c d}} F_{\underline b \overline {c d}} \, ,
\ee
where we have defined
\be
D_{\overline a} = E^M{}_{\overline a} \partial_M \ , \ \ \ D_{\underline a} = E^M{}_{\underline a} \partial_M  \ ,
\ee
and
\begin{subequations}
\begin{align}
F_{\overline a \underline {b c}} &= D_{\overline a} E_M{}_{[\underline b} E^M{}_{\underline{c}]} + 2 D_{[\underline b} E^M{}_{\underline c]} E_{M \overline a} \, ,\\\
F_{\underline a \overline {b c}} &= D_{\underline a} E_M{}_{[\overline b} E^M{}_{\overline{c}]} + 2 D_{[\overline b} E^M{}_{\overline c]} E_{M \underline a}\, .
\end{align}
\end{subequations}
The parameters $a$ and $b$ interpolate between generalized Green-Schwarz transformations with respect to the two different Lorentz factors. They were originally found in \cite{hohmzwiebach}, and give rise to a two-parameter family of theories which contains the bosonic string $(a,b)=(-\alpha',-\alpha')$, the heterotic string  $(a,b)=-(\alpha',0)$ and more general deformations like those in \cite{hsz}. Interestingly, to first order in $\alpha'$ DFT is defined through a Lagrangian that is invariant under all these symmetries \cite{marn}, and transforms as a scalar with respect to generalized diffeomorphisms. This property will be inherited by the supergravity Lagrangian that descends from it, and also by all others connected by field redefinitions.

Making contact with supergravity degrees of freedom requires performing a $GL(D)$ decomposition of $O(D,D)$, with respect to which the indices split as $M = ({}^\mu, \, {}_\mu)$. Selecting  the supergravity section as a solution of the strong constraint, simply amounts to setting $\partial_M = (0 ,\, \partial_\mu)$.
The decomposition of the invariant metrics is given by
\be
\eta_{M N} = \left(\begin{matrix} 0 & \delta^\mu_\nu  \\ \delta^\nu_\mu & 0\end{matrix} \right) \ , \ \ \ \eta_{\underline {a b}} = - g_{\underline {a b}} \ , \ \ \ \eta_{\overline {ab}} = g_{\overline {ab}} \, ,
\ee
where $g_{\underline {a b}}$ and $g_{\overline {a b}}$ are Minkowski metrics.  The most general $GL(D)$ covariant parameterization of the generalized fields is 
\be
{E}_{M}{}^{\underline a}=\frac1{\sqrt2}\left(\begin{matrix} \bar e^\mu{}_{\underline c}g^{\underline{ca}} \\
(\bar b_{\mu\nu}-\bar g_{\mu\nu})\bar e^{\nu}{}_{\underline c}g^{\underline{ca}} \end{matrix}\right)  \ , \ \ \ {E}_{M}{}^{\overline a}=\frac1{\sqrt2}\left(\begin{matrix}  \bar e^\mu{}_{\overline c}g^{\overline{ca}} \\ (\bar g_{\mu\nu}+\bar b_{\mu\nu})\bar e^{\nu}{}_{\overline c}g^{\overline{ca}}
\end{matrix}\right) \ , \ \ \ e^{-2d}=\sqrt{-\bar g}e^{-2\bar\phi} \, , \nonumber
\ee
where we have introduced a two-form $\bar b_{\mu \nu}$, a dilaton $\bar \phi$ and a pair of vielbeins $\bar e_\mu{}^{\underline a}$ and $\bar e_\mu{}^{\overline a}$  that generate the same metric $\bar g_{\mu \nu}$
\be \bar g_{\mu \nu} = \bar e_\mu{}^{\underline a} g_{\underline {a b}} \bar e_\nu{}^{\underline b} = \bar e_\mu{}^{\overline a} g_{\overline {a b}} \bar e_\nu{}^{\overline b}\, ,\ee
which raises and lowers the $GL(D)$ indices.  Finally, the element in the algebra of $O(D,D)$ splits into constant components
\ba
h_M{}^N=\left(\begin{matrix}a^\mu{}_\nu&\beta^{\mu\nu} \\
B_{\mu\nu}&-a^\nu{}_\mu\end{matrix}\right)\, ,
\ea
where $a^\mu{}_\nu$ generates rigid $GL(D)$ rotations, $B_{\mu\nu}$ produces rigid shifts of the two-form and $\beta^{\mu\nu}$ is a bi-vector that mixes the gravitational and two-form sector. In this paper we will consider $a^\mu{}_\nu = B_{\mu \nu} = 0$ because their action in supergravity is trivial, and focus only on $\beta$ transformations. Notice that in DFT the derivatives belong to the fundamental representation of $O(D,D)$, and so transform linearly under $O(D,D)$. In particular, the $\beta$ elements break the choice of section 
\be
\delta_\beta \partial_M \dots = h_M{}^N \partial_N  \dots = (\beta^{\mu \nu} \partial_\nu , \, 0) \dots\, ,
\ee
which must be restored by imposing the constraint
\be
\beta^{\mu \nu} \partial_\nu \dots = 0\, . \label{betaconstraint}
\ee

To establish a connection with supergravity, we must break
the double Lorentz symmetry to a single Lorentz transformation and gauge fix the two vielbeins to a single one 
\be
\bar e_\mu{}^a = \bar e_\mu{}^{\underline a} \delta^a_{\underline a} = \bar e_\mu{}^{\overline a} \delta^a_{\overline a} \, .
\ee
To this end we have introduced Kronecker deltas to force the two Lorentz groups to carry the
same set of indices $a, b, c, \dots$, which will be the standard Lorentz indices in supergravity. The gauge fixing applied to the flat derivatives yields
\be
D_{\underline a} = - \frac 1 {\sqrt{2}} \delta^a_{\underline a} \, D_a  \ , \ \ \ \  D_{\overline a} = \frac 1 {\sqrt{2}} \delta^a_{\overline a} \, D_a \ , \ \ \ \ D_a = \bar e^\mu{}_a \partial_\mu \ ,
\ee
and to the generalized fluxes (see Appendix \ref{AppA} for definitions.)
\ba
F_{\overline {a} \underline{b c}} &=& \frac 1  {\sqrt{2}}  \delta^{a}_{\overline a} \delta^b_{\underline b} \delta^c_{\underline c} \left(w_{abc} - \frac 1 2 H_{abc}\right) \, , \\ 
F_{\underline {a} \overline{b c}} &=& \frac 1  {\sqrt{2}}  \delta^{a}_{\underline a} \delta^b_{\overline b} \delta^c_{\overline c} \left(w_{abc} + \frac 1 2 H_{abc}\right) \, . 
\ea

Because we have gauge  fixed the double Lorentz symmetry to its diagonal subgroup, the two Lorentz parameters are no longer independent. We must then explore how they are related, and what is the most convenient way to express them in terms of the Lorentz parameter in supergravity.
To achieve this, we write the two Lorentz invariant metrics in terms of a single one
\be
g_{a b} = \delta_a^{\underline a} \delta_b^{\underline b} g_{\underline {a b}} = \delta_a^{\overline a} \delta_b^{\overline b} g_{\overline {a b}} \, ,
\ee
and also express all the Lorentz parameters (including the generalized Green-Schwarz deformation $\Delta$) in terms of the same set of indices
\be
\Lambda_{\overline {ab}} = \delta^a_{\overline a} \delta^b_{\overline b} \overline \Lambda_{ab}  \ , \ \ \ \ \Lambda_{\underline {ab}} = \delta^a_{\underline a} \delta^b_{\underline b} \underline \Lambda_{ab} \ , \ \ \ \
\Delta^{(1)}_{\overline {a} \underline b} = \delta^a_{\overline a} \delta^b_{\underline b} \Delta^{(1)}_{ab} \ . \label{GaugeFixLorentzParam}
\ee
In this way, we end up with two different transformations for the gauge fixed vielbein, each one coming from the transformation of each component of the generalized frame
\begin{subequations}
\begin{align}
\delta E^{\mu \underline a} \ \ \ \ \ \to \ \ \ \ \ \delta \bar e^\mu{}_{a} \ &= \  \bar e^{\mu b} \left( \beta_b{}^c \bar b_{c a} - \beta_{b a} -\underline \Lambda_{b a} - \Delta^{(1)}_{b a}\right)  \ , \\
\delta E^{\mu \overline a} \ \ \ \ \ \to \ \ \ \ \ \delta \bar e^\mu{}_{a} \ &= \ \bar e^{\mu b} \left(\beta_b{}^c \bar b_{c a} + \beta_{ba}+ \overline \Lambda_{b a} - \Delta^{(1)}_{a b}\right) \ , \label{transfoverlinebien}
\end{align}
\end{subequations}
where
\be
\Delta^{(1)}_{ab}=- \frac a4D_{ b}\underline\Lambda^{{cd}} \left(w_{acd} - \frac 1 2 H_{a c d} \right) + \frac b4D_{ a}\overline\Lambda^{{cd}}\left(w_{bcd} + \frac 1 2 H_{b c d} \right) \label{Delta1}\, .
\ee
These transformations must obviously coincide, which imposes the required relation between the double Lorentz parameters
\be
\underline \Lambda_{a b} = - \overline \Lambda_{a b} - 2 \Delta^{(1)}_{[ab]} - 2\beta_{a b} \ . \label{gf}
\ee
Up to redefinitions of the standard Lorentz parameter of supergravity $\Lambda_{a b}$, this equation is parametrically solved  as
\be
\underline\Lambda_{ab} = -\Lambda_{ab}-\beta_{ab}-\Delta^{(1)}_{[ab]} \ , \ \ \
\overline\Lambda_{ab}=\Lambda_{ab}-\beta_{ab}-\Delta^{(1)}_{[ab]}\, .
\ee
Combining these expressions with (\ref{Delta1})  and keeping only terms up to first order in $\alpha'$, we find that  Lorentz and $\beta$ transformations of the gauge fixed fields in the DFT scheme of supergravity read
\begin{subequations}
\begin{align}
\delta \bar e_{\mu a} \ &= \ - \bar e_\mu{}^b \left(\beta_a{}^c \bar b_{ c b} + \Lambda_{a b} - \Delta^{(1)}_{(a b)}\right)\, ,\\
\delta\bar\phi \ &= \ \frac12\left(\beta^{\mu \nu} \bar b_{\mu \nu}+ \Delta^{(1)}_a{}^a\right)\, , \\
\delta \bar b_{\mu \nu} \ &= \ -\beta_{\mu\nu}-\bar b_{\mu\rho}\beta^{\rho\sigma}\bar b_{\sigma\nu}- 2\bar e_\mu{}^a\bar e_\nu{}^b\Delta^{(1)}_{[ab]}\, ,
\end{align}
\end{subequations}
where
\ba 
\Delta^{(1)}_{ab}&\equiv&\frac {a+b}4\left(D_{(a}\Lambda^{{cd}}w_{b)cd}-\frac12D_{(a}\beta^{{cd}}H_{b)cd}\right)+\frac {b-a}4\left(\frac12D_{ (a}\Lambda^{{cd}}H_{b)cd}
-D_{ (a}\beta^{{cd}}w_{b)cd}\right)\, \nn\\
&+&  \frac {a+b}4\left(\frac12 D_{ [a}\Lambda^{{cd}}H_{b]cd} - D_{[a }\beta^{{cd}}w_{b]cd}\right)+\frac {b-a}4\left(D_{ [a}\Lambda^{{cd}}w_{b]cd}
+\frac12 D_{[a}\beta^{{cd}}H_{b]cd}\right)\, . \ \ \ \ \ \ \
\ea

In summary, we have derived the first order deformations of Lorentz and $\beta$ transformations in the DFT scheme of supergravity, namely the scheme in which the fields are given by a gauge fixing of the duality covariant components of the generalized fields in DFT. 

It is instructive to cast the result in the form given in \cite{bmn}, which allows to find purely flattened expressions for the transformation rules. Defining\footnote{It is important to note that, contrary to the conventions in the rest of the paper, here $\delta b_{ab}$ is the flat version of $\delta b_{\mu \nu}$ and {\it not} the variation of $b_{ab}$.}
\be
\delta \bar e_{a b} = \bar e^\mu{}_a \, \delta \bar e_{\mu b} \ ,  \ \ \ \   \delta \bar \phi = \frac 1 2 \delta \bar e_a{}^a \ , \ \ \ \ \delta \bar b_{a b} = \bar e^\mu{}_a \bar e^\nu{}_b \, \delta \bar b_{\mu \nu} \ , \label{defdeltas}
\ee 
the leading order transformation rules are
\begin{subequations} \label{transf0}
\begin{align}
\delta^{(0)} \bar e_{a b} &= - \bar b_a{}^{c} \beta_{c b} + \Lambda_{a b}\, ,\\ 
\delta^{(0)} \bar b_{a b} &=  - \beta_{a b} - \bar b_{a c} \beta^{c d} \bar b_{d b} \, ,
\end{align}
\end{subequations}
and the first order $\alpha'$-corrections are
\begin{subequations} \label{transfL1}
\begin{align}
\delta^{(1)}_\Lambda \bar e_{a b} &= \frac {a + b} 4 D_{(a}\Lambda^{cd} w_{b) c d} + \frac{b-a} 8 D_{(a}\Lambda^{c d} H_{b)cd}\, ,\\
\delta^{(1)}_\Lambda \bar b_{a b} &= - \frac {a + b } 4 D_{[a}\Lambda^{cd}H_{b]cd} - \frac {b-a} 2 D_{[a} \Lambda^{cd} w_{b] c d}\, , \label{BosonicLorentz}
\end{align}
\end{subequations}
and
\begin{subequations} \label{transfbeta1}
\begin{align}
\delta^{(1)}_\beta \bar e_{a b} &= - \frac {a + b} 8 D_{(a}\beta^{c d} H_{b) c d} - \frac {b-a} 4 D_{(a}\beta^{c d} w_{b)cd}\, ,\\
\delta^{(1)}_\beta \bar b_{a b} &= \frac {a+ b} 8 D_{[a} \beta^{cd} w_{b]cd} - \frac {b-a} 4 D_{[a} \beta^{c d} H_{b]cd}\, .
\end{align}
\end{subequations}

These transformations can be cast in other schemes, performing field redefinitions $ \{\bar e,\, \bar \phi,\, \bar b\} \to \{e,\, \phi, \, b \}$ of the form
\begin{subequations}\label{fg}
\begin{align}
\bar e_\mu{}^a \ &= \  e_\mu{}^b (\delta_b^a + F^a{}_b) \ , \ \ \ e_\mu{}^a \ = \ \bar e_\mu{}^b (\delta_b^a - F^a{}_b)\, ,\\
\bar \phi \ &=\  \phi + \frac 1 2 F_a{}^a \ , \ \ \ \ \ \ \ \ \ \bar b_{\mu \nu}\ =\ b_{\mu \nu} - G_{\mu \nu}\, ,
\end{align}
\end{subequations}
where both $F_{ab}$ and $G_{\mu \nu}$  start at first order in $\alpha'$. Therefore, for the purpose of obtaining the first order deformations of the transformation rules, these functions can be expressed in terms of either the new (unbarred)  or the previous (barred) set of fields in the DFT scheme. Since we are only interested in connecting with schemes that preserve the generalized measure $\sqrt{-\bar g} e^{-2\bar \phi} = \sqrt{-g} e^{-2\phi}$, the redefinition of the dilaton is determined by that of the vielbein.

The functions $F_{a b}$ and $G_{\mu \nu}$ define the scheme, and the Lorentz and $\beta$ transformations in different schemes are given, up to redefinitions of the parameters,  by 
\begin{subequations}\label{tfg}
\begin{align}
\delta e_{a b} \ &= \ - (b_a{}^c - G_a{}^c) \beta_{c b} + \Lambda_{ab} + \Delta^{(1)}_{(ab)} - \Delta F_{b a} \, ,\\
\delta b_{a b}\ &= \ - \beta_{a b} - b_{ac} \beta^{cd} b_{db} - 2 \Delta^{(1)}_{[a b]} - 4 \beta_{[a}{}^c  F_{(b] c)} + 2 \beta^{c d} b_{c [a} G_{b]d} + e^\mu{}_a e^\nu{}_b \delta G_{\mu \nu} \, ,\ \ \ \  \ \ \ \ \ \\
\delta \phi\ &=\ \frac 1 2 \delta e_a{}^a\, ,
\end{align}
\end{subequations}
where all the fields  belong to the new (unbarred) scheme, and 
\be
\Delta F_{a b} = \delta F_{a b} - \Lambda^c{}_a F_{c b} - \Lambda^c{}_b F_{a c} \ ,
 \ee
denotes the non-covariant transformation of the field redefinition.

In forthcoming sections  we will specify the functions $F_{a b}$ and $G_{\mu \nu}$ that connect the DFT scheme with those of Bergshoeff-de Roo (BdR) and Metsaev-Tseytlin (MT).

\section{Scheme independent statements}

Although the first order corrections to the $\beta$ transformations depend on the scheme, there are important issues that are independent of field redefinitions. In this section we discuss some of these scheme independent questions:   $\beta$ symmetry at leading order, redundancies in the $\beta$ transformation rules at ${\cal O}(\alpha')$, the relation between closure of the symmetry algebra and the invariance of the action/Lagrangian, and the effect of field and parameter redefinitions in the closure.

\subsection{$\beta$ symmetry at leading order}

The two-derivative DFT Lagrangian descends to the universal two derivative NS-NS Lagrangian 
\be
L^{(0)}= R - 4 (\nabla \phi)^2  + 4 \Box \phi - \frac 1 {12} H^2 \ , \label{DSugra}
\ee
which must then be invariant under the transformations \eqref{transf0}, 
\be
\delta_\beta e_{a b}= - b_{a c} \beta^c{}_b \ , \ \ \ 
\delta_\beta b_{a b}= - \beta_{a b} - b_{a c} \beta^{c d} b_{d b}  \ , \ \ \ 
\delta_\beta \phi =  \frac 1 2 \delta e_a{}^a  \ , \label{betatransf}
\ee
whenever the constraint \eqref{betaconstraint} holds, namely
\ba
\beta^{\mu\nu}\partial_\nu\cdots =0\, .\label{const}
\ea 
Using that $\beta^{\mu \nu}$ is constant\footnote{Here, $\beta$ is constant  because we are dealing with tori. We thank E. \'O Colg\'ain for pointing out that a non-constant $\beta$ could be connected to Yang-Baxter  deformations.} and antisymmetric, the following useful identities are obtained 
\ba
\nabla_a\beta^{bc}=2w_{da}{}^{[b} \beta^{c]d}\, , \qquad \beta^{ab}w_{abc}=0\, , \qquad
\omega_{[ab]}{}^{c} \beta_1^{ad} \beta_2^{be}=0\, , \label{identities}
\ea
where the last one is particularly useful for the study of the symmetry algebra. 
It is instructive to compute the $\beta$ transformations of many of the tensors and connections that appear in supergravity 
\begin{subequations} \label{transftensors}
%\begin{multline}
\begin{align}
\delta_\beta \left( \sqrt{- g} e^{- 2 \phi} \right)\ &= \ 0 \, , \\
\delta_\beta w_{cab}\ &=\ \beta_{[a}{}^dH_{b]cd}-\frac12\beta_c{}^dH_{abd}  \, ,\\
\delta_\beta H_{abc}\ &=\ - 3 \nabla_{[a}\beta_{b c]}  \, ,\\
\delta_\beta(D_a\phi)\ &=\ \frac12\beta^{cd}H_{acd}  \, ,\\
\delta_\beta( \nabla_a \nabla_b\phi) \ &=\ \frac12\nabla_{(a}\left(\beta^{cd} H_{b)cd}\right) - \beta^{c}{}_{(a} H_{b) c d} \nabla^{d}\phi  \, , \\
\delta_{\beta} R_{a b}{}^{ c d}\ &= \
- \nabla_{[a} \left(\beta_{b]}{}^{e} H_{e}{}^{c d}\right)  - \nabla^{[c} \left(\beta^{d]e} H_{e a b}\right) 
\nonumber  \, ,\\ 
& \ \ \ \ +\left( \nabla_{[a}\beta^{[ce} -  \nabla^{[c}\beta_{[a}{}^{e} -  \nabla^{e}\beta_{[ a}{}^{[c}   \right) H^{d]}{}_{ b]e}  \, ,\\
\delta_\beta R_{ab}\ &=\ - \nabla^{c} \left( \beta^{d}{}_{(a}  H_{b) c d} \right) 
- \nabla_{(a}\left(\beta^{cd} H_{b)cd}\right) - \frac12 \nabla_{(a}\beta^{cd} H_{b) c d}\, ,\\
\delta_\beta R \ &=\ - 2\nabla^a\left( \beta^{b c}  H_{a b c}\right)-\frac12 \nabla^{a} \beta^{b c} H_{a b c}\;.
\end{align}
%\end{multline}
\end{subequations} 
These identities can now be used to fix the couplings of a generic combination of gauge invariant terms, by requiring $\beta$ symmetry. Indeed,
\begin{eqnarray}
0&=& \delta_\beta\left[\sqrt{-g} f(\phi)\left(R + c\, \Box \phi + d \left(\nabla\phi\right)^2+ e\, H^{2}\right)\right]\nn\\
&=&
 \sqrt{-g} f(\phi)\left[ (c-4) \nabla^{a}\left(\beta^{b c} H_{a b c}\right)   - (12  e + 1) \nabla^{a} \beta^{b c} H_{a b c} + 2 (c+d) \nabla^{a}\phi \beta^{b c} H_{a b c}\right]\cr\cr
 &&+ \frac12 \sqrt{-g}   \beta^{ab} b_{ab}  \left(\vphantom{2^{\frac12}} 2 f(\phi) + f'(\phi) \right)  \;\left[R + c\, \Box \phi + d \left(\nabla\phi\right)^2+ e\, H^{2}\right]    \;  ,
\end{eqnarray}
determines  $c=-d=4$, $e=-\frac{1}{12}$ and $f(\phi)=e^{-2\phi}$,  in perfect agreement with supergravity \eqref{DSugra}. 

Note that $\beta$ transformations realize the non-geometric sector of $O(d, d)$ as a hidden symmetry in
the standard supergravity scheme. Instead, they become both geometric and manifest in the
so-called $\beta$-supergravity scheme \cite{andriot}, where the non-geometric sector is realized by the B-shifts,
which should then fix the corresponding couplings. 

Finally, we notice that $\beta$ and Lorentz transformation rules close off-shell to lowest order as follows
\be
\left[\delta_1\, , \, \delta_2\right] = - \delta_{12} \ , \ \ \ \Lambda_{1 2 a b} = 2 \, \beta_{1[a}{}^{c} \beta_{2\, b] c} 
    + 2\, \Lambda_{1 [a}{}^{c} \Lambda_{2\, b] c} \ . \label{closure}
\ee

\subsection{A four-parameter redundancy of first order transformations} \label{4param}

When the variation of the dilaton leaves the measure $\sqrt{-g}e^{-2\phi}$ invariant, i.e. $\delta \phi = \frac 1 2 \delta e_a{}^a$, the variation of $L^{(0)}$  can be decomposed as
\be
\delta{ L}^{(0)}= {\cal E}{}^{ab} \; \delta  e_{ab}  + {\cal B}^{ab} \; \delta b_{ab} 
+ {\cal D}_{a}\left[2\; {\cal D}_{b}\delta e^{ab} -\frac12 H^{abc} \delta b_{bc}\right]\, .\label{d1L0}
\ee
Here we defined the operator
\be
{\cal D}_a X = {e^{2 \phi}}\nabla_a \left(e^{-2\phi} X \right) = \nabla_{a} X -2\nabla_{a}\phi X  \ , \label{calD}
\ee
acting on a generic tensor $X$. Although it is not formally a derivative as it does not obey the Leibniz rule, it leads to a total derivative  when inserted into the action, and so can be discarded in such context. Additionally, we introduced the tensors that define the equations of motion of the frame ${\cal E}_{ab}$ and the two-form ${\cal B}_{ab}$
\ba
{\cal E}_{ab} &=& - 2 \left(\vphantom{\frac12}  R_{a b} + 2 \nabla_{a}\nabla_{b}\phi -\frac14 H_{a c d} H_{b}{}^{c d} \right)\, , \label{eome} \\
{\cal B}_{a b} &=& \frac12 {\cal D}_{c}H_{ab}{}^c{} \label{eomb} \ .
\ea

A useful observation is that, if  the total derivative term in (\ref{d1L0}) is ignored, namely if one focuses on  invariance of the action instead of the Lagrangian, $\delta L^{(0)}$ (\ref{d1L0}) vanishes  up to total derivatives for  variations of the form
\ba
\delta e_{ab} &=& T^1_{[(ab)(cd)]} {\cal E}^{cd} + T^2_{(ab)[cd]} {\cal B}^{ c d} \ ,\\
\delta b_{ab} &=& - T^2_{(cd)[ab]} {\cal E}^{ cd} + T^3_{[[ab][cd]]} {\cal B}^{ c d} \ .
\ea
The  $T$ tensors can depend on anything as long as their symmetries are respected. In particular, if they are linear in the  parameters of the transformation, and contain no derivatives,  they account for first order deformations that leave the action invariant. 

For $\beta$ transformations, there are only four possible deformations of this type  to first order in $\alpha'$, namely 
\begin{subequations}
\begin{align}
T^1_{ab}{}^{cd} &= \kappa \beta_{(a}{}^{(c} g_{b)}{}^{d)}\, , \\
T^2_{ab}{}^{cd} &= \gamma  \beta_{(a}{}^{[c} g_{b)}{}^{d]} +  \chi g_{ab} \beta^{cd}\, ,  \\
T^3_{ab}{}^{cd} &= \rho \beta_{[a}{}^{[c} g_{b]}{}^{d]} \, ,
\end{align}
\end{subequations}
which in turn define the following four-parameter ($\gamma, \chi, \kappa, \rho$) family of first order transformations 
\begin{subequations}\label{deformparam2}
\begin{align}
\delta_\beta^{(1)} e_{a b} &= \gamma \beta_{c(a} {\cal B}_{b)}{}^{ c} +  \chi g_{a b} \beta_{c d} {\cal B}^{ c d} + \kappa  \beta_{(a}{}^c {\cal E}_{b)c}\, , \\
\delta_\beta^{(1)} b_{a b} &= - \gamma \beta^c{}_{[a} {\cal E}_{b]c}  -  \chi \beta_{a b} {\cal E}_c{}^c + \rho \beta_{c[a} {\cal B}^{c}{}_{b]}\, . 
\end{align}
\end{subequations}
This result is scheme independent, as it simply amounts to a first order invariance of the lowest order action. Then, in any scheme, adding these first order deformations to the  $\beta$ transformations, respects the invariance of the action to first order, for whatever values of ($\gamma, \chi, \kappa, \rho$). Let us emphasize that, when the transformations are deformed by these terms, the invariance of the Lagrangian is only achieved up to total derivatives. 

Besides these, there are still extra symmetries of the two-derivative action, of the form  
\begin{subequations}\label{deformreparam2}
\begin{align}
\delta_\beta^{(1)} e_{a b} =& f_{[ab]}(\beta,\psi_i) +\nabla_{(a}h_{b)}(\beta,\psi_i)\, , \\
\delta_\beta^{(1)} b_{a b} =& \nabla_{[a}t_{b]}(\beta,\psi_i) + H_{abc} h^{c}(\beta,\psi_i)\, ,
\end{align}
\end{subequations}
where $f,h$ and $t$ are arbitrary functions, that can be interpreted as due to $\beta$ and $\psi_i$ field dependent reparameterizations of Lorentz, diffeomorphisms and gauge parameters. It is worth noticing that, while $f$ and $t$ describe an exact symmetry of the Lagrangian, $h$ is in general only a symmetry of the action. Exceptions can occur, for instance when $h_{a}\sim\beta_{ab} X^{b}$ for arbitrary functions $X^b$, because diffeomorphisms with parameter $\xi^{\mu}=\beta^{\mu\nu} X_{\nu}$ leave the Lagrangian exactly invariant. In fact in the conventional KK reduction, the parameter $\xi^{\mu}=\beta^{\mu\nu} X_{\nu}$ has only internal components and so represents a gauge transformation, which leaves the Lagrangian invariant. We will go back to the role of reparameterizations in subsection \ref{BracketRedef}. 

In conclusion, any first order deformation of the $\beta$ transformation rules that leave the four-derivative action invariant, is determined up to the four parameters $\gamma,\kappa,\chi,\rho$ and three functions $f,h,t$.

We would like to emphasize that the deformations \eqref{deformparam2} cannot be eliminated (either completely or partially) by field and/or parameter redefinitions. This readily follows from the observation that the vanishing of the total derivative in \eqref{d1L0}  is a necessary condition for the existence of field and/or parameter redefinitions that cancel \eqref{deformparam2}.

Indeed, suppose we start with a generic transformation depending on all the parameters $a, b, \gamma, \chi, \rho, \kappa$, and some choice of the last four could be absorbed by a field and/or parameter redefinition. In the scheme defined by that particular choice, the symmetry transformations would reduce to \eqref{tfg} for some unknown F and G. Since \eqref{tfg}  are related to \eqref{transfbeta1} by field redefinitions and \eqref{transfbeta1} preserve the Lagrangian exactly,  we conclude that there is no such choice if the original Lagrangian was not exactly invariant.  In fact,  the invariance of the Lagrangian cannot be modified by field and/or parameter redefinitions.

Moreover, it is relatively straightforward to verify that no non-trivial choice of parameters can cancel the total derivative. Actually, the problem can be simplified as $\gamma, \chi$ ($\kappa,\rho$) lead to odd (even) powers of the field strength $H$, so they must cancel separately.

\subsection{Invariance of the action implies closure of the symmetry algebra} \label{InvarianceClosure}

So far we have found the  symmetries that leave the first order Lagrangian (in the DFT scheme) invariant. Now we point out that invariance of the Lagrangian with respect to a complete set of symmetries implies that the algebra of transformations closes off-shell. Later, we will show that the deformations discussed in the previous section, i.e. transformations that depend on the equations of motion and leave the action  (instead of the Lagrangian) invariant, still respect closure, but only on-shell. Again, the contents of this section are scheme independent. Similar results along these lines can be found in \cite{hohmbonezzi}.

 The algebra of transformations that leave a Lagrangian invariant is expected to close on general grounds. To see this, consider a generic Lagrangian  depending on fields $\psi_i$ and their derivatives up to an arbitrary order, namely $L(\psi_i,\partial_{\mu}\psi_i,\partial_{\mu}\partial_{\nu}\psi_i\, ,...)$. We assume that terms with derivatives of different orders are independent. If $L$  is invariant under a complete set of local and global transformations, infinitesimally parameterized by   $\zeta_A$, then
\begin{eqnarray}
0=\delta_{\zeta} L = \sum_{n = 0}^\infty \frac{\partial  L}{\partial (\partial^n \psi_i)} \partial^n \delta_\zeta \psi_i= \frac{\partial L}{\partial\psi_i} \delta_{\zeta}\psi_i  
+ \frac{\partial L}{\partial(\partial_{\mu}\psi_i)} \partial_{\mu}\delta_{\zeta}\psi_i
+ \frac{\partial L}{\partial(\partial_{\mu}\partial_{\nu}\psi_i)} \, \partial_{\mu}\partial_{\nu} \delta_{\zeta}\psi_i + ...\, , \ \ \ \ \label{ifonlyif}
\end{eqnarray}
where the dots stand for contributions depending on higher-order derivatives. The above sum vanishes {\it if and only if} the variation is performed with respect to a symmetry transformation. 

Now we consider the commutator of two variations
\ba
0 = [\delta_{\zeta_1} ,\delta_{\zeta_2}] L &=& \sum_{n = 0}^\infty \frac{\partial  L}{\partial (\partial^n \psi_i)} \partial^n \left(\vphantom{e^{\frac12}} [\delta_{\zeta_1} , \delta_{\zeta_2}] \psi_i \right) \label{brL}\\ && +  2 \sum_{n = 0}^\infty \sum_{m = 0}^\infty \frac{\partial^2  L}{\partial (\partial^n \psi_i)\partial (\partial^m \psi_j)} \left(\partial^n \delta_{\zeta_{[1}} \psi_i \right) \left(\partial^m \delta_{\zeta_{2]}} \psi_j \right) \ . \nonumber\ea 
The last term here cancels due to the symmetry under the exchange $(n,i) \leftrightarrow (m ,j)$ and the antisymmetry with respect to $1 \leftrightarrow 2$.
Comparing the first term with (\ref{ifonlyif}), we conclude that $[\delta_{\zeta_1} ,\delta_{\zeta_2}] L$ vanishes {\it if and only if} $[\delta_{\zeta_1} ,\delta_{\zeta_2}] \psi_i $ amounts to a symmetry transformation, namely
\be
[\delta_{\zeta_1} ,\delta_{\zeta_2}] \psi_i = - \delta_{\zeta_{12}} \psi_i \ , \label{closureoffshell}
\ee
for some parameter (bracket) $\zeta_{1 2}$. Since we only assumed that the Lagrangian is invariant, we conclude that invariance of the Lagrangian implies that the symmetry algebra closes off-shell.

Now we deform the transformations that leave the Lagrangian invariant as discussed in the previous subsection, i.e.  we consider the deformations (\ref{deformparam2}) with parameters ($\gamma, \chi, \kappa, \rho$) that are first order in $\alpha'$. Then, the LHS of (\ref{closureoffshell}) becomes, to first order,
\ba
[\delta_{\zeta_1} + \delta_{\zeta_1}^{(\gamma, \chi,
\kappa, \rho)}, \delta_{\zeta_2} + \delta_{\zeta_2}^{(\gamma, \chi,
\kappa, \rho)}] \psi_i = [\delta_{\zeta_1} , \delta_{\zeta_2}] \psi_i + 2 \delta^{(0)}_{\zeta_{[1}} \delta_{\zeta_{2]}}^{(\gamma, \chi,
\kappa, \rho)} \psi_i + 2 \delta_{\zeta_{[1}}^{(\gamma, \chi,
\kappa, \rho)} \delta^{(0)}_{\zeta_{2]}} \psi_i\, . \label{deformedclosure}
\ea
 Here, the first term in the RHS gives the bracket $\delta_{\zeta_{12}} \psi_i$ discussed above. The last term vanishes on-shell because the ($\gamma, \chi,
\kappa, \rho$) variations depend on the lowest order equations of motion. The second term is more subtle, because it involves variations of the equations of motion. Under the standard symmetries they transform covariantly, but we have to make sure that the same is true for the lowest order $\beta$ transformations. Indeed, we find that $\beta$ transformations mix the equations of motion of the vielbein and two-form as follows
 \be
 \delta_\beta^{(0)} {\cal E}_{a b} = -4 \beta_{c(a} {\cal B}_{b)}{}^{c} \ , \ \ \  \delta_\beta^{(0)} {\cal B}_{a b} = \beta_{c[a} {\cal E}_{b]}{}^{ c}\, ,
 \ee
which then implies that the second term in the RHS of  (\ref{deformedclosure}) also  vanishes on-shell. In summary, the deformed transformations that leave the action  invariant, close on-shell
\ba
[\delta_{\zeta_1} + \delta_{\zeta_1}^{(\gamma, \chi,
\kappa, \rho)}, \delta_{\zeta_2} + \delta_{\zeta_2}^{(\gamma, \chi,
\kappa, \rho)}] \psi_i = -(\delta_{\zeta_{12}} + \delta_{\zeta_{12}}^{(\gamma, \chi,
\kappa, \rho)}) \psi_i + \dots \, , \label{closuredeformed}
\ea
where the dots represent terms proportional to the equations of motion, and we have the freedom to add the last term in (\ref{closuredeformed}) as it also vanishes on-shell.

\subsection{Effect of field and parameter redefinitions on the symmetry algebra} \label{BracketRedef}
In section \ref{dft} we examined the effect of field redefinitions on the transformation rules, when moving from one scheme to another. Here we further explore the role of field and parameter redefinitions on the bracket. 

Even though field redefinitions preserve the closure of the symmetry algebra, the precise form of the bracket could be altered.  Indeed, consider fields $\psi_{i}$ satisfying
\begin{eqnarray}
\left[\delta_{\zeta_1},\delta_{\zeta_2}\right]\psi_{i}=-\delta_{\zeta_{12}(\psi)}\psi_{i}\, ,
\end{eqnarray} 
where we have made explicit that the bracket $\zeta_{12}(\psi)$ could eventually be field dependent. Let us consider the field redefinition
\begin{eqnarray}
\psi_{i}\to\psi'{}_{i}=\psi{}_{i}+\Delta \psi{}_{i} = \psi{}_{i} + \sum a_{i}{}^{j_1 j_2 ...}\partial^{n_1}\psi_{j_1}\partial^{n_2}\psi_{j_2} \dots
\label{psiredef}
\end{eqnarray}
Here $\partial^{n}$ denotes derivatives of order $n=0,1,2,...$ and ellipsis stand for extra field dependence. So, it parameterizes a generic field redefinition with no restriction on the number of fields and derivatives thereof. The bracket transforms the new field as
\begin{eqnarray}
\left[\delta_{\zeta_1},\delta_{\zeta_2}\right] \psi'{}_{i} 
 &=& 
\left[\delta_{\zeta_1},\delta_{\zeta_2}\right]\psi{}_{i}  
+ \sum a_{i}{}^{j_1 j_2 ...}\partial^{n_1}\left(\left[\delta_{\zeta_1},\delta_{\zeta_2}\right]\psi_{j_1}\right)\partial^{n_2}\psi_{j_2} \dots \cr
&&+\
\sum a_{i}{}^{j_1 j_2 ...}\partial^{n_1}\psi_{j_1}\partial^{n_2}\left(\left[\delta_{\zeta_1},\delta_{\zeta_2}\right]\psi_{j_2}\right) \dots
+\dots \cr
&=& -\delta_{\zeta_{12}(\psi)}\psi'_{i}\;,
\end{eqnarray}
where in the first equality we used a symmetry argument similar to the one followed in (\ref{brL}), implying that terms where $\delta_{\zeta_1}$ and $\delta_{\zeta_2}$ act on different fields cancel each other after antisymmetrization of indices $(1\leftrightarrow 2)$.

Plugging the derivative expansion of the old bracket at ${\cal O} (\alpha')$
\begin{eqnarray}
\zeta_{12}(\psi)=\zeta^{(0)}_{12}(\psi)+\zeta^{(1)}_{12}(\psi) \; , 
\end{eqnarray}
into (\ref{psiredef}), we get the new bracket
\begin{eqnarray}
\left[\delta_{\zeta_1},\delta_{\zeta_2}\right] \psi'{}_{i} 
 = -\delta_{\zeta'_{12}(\psi')}\psi'_{i}\;\;,\ \ \ \ \ \ \ \ \ \ 
\zeta'_{12}(\psi'):= \zeta_{12}(\psi')-\frac{\partial \zeta^{(0)}_{12}}{\partial \psi_i} \Delta \psi_i\;,\label{BraFred}      
\end{eqnarray}
where we have considered that $\zeta^{(0)}_{12}$ eventually depends on the fields $\psi_i$, but not on derivatives thereof.
It is worth noticing here that in general the bracket gets no corrections after a field redefinition, because the leading order bracket $\zeta^{(0)}_{12}$ is in general field independent. Although this is the case for diffeomorphisms, Lorentz and gauge transformations, for  $\beta$ transformations it is not (see Appendix \ref{AppB}).

%\bigskip
Redefinitions of the symmetry parameters are more subtle.
Sometimes it is convenient to redefine them so that the transformation rules look more suitable. For instance, an antisymmetric piece in $\delta^{(1)}_{\beta}e_{ab}$ can be canceled with an appropriate choice of the function $f_{[ab]}$ in (\ref{deformreparam2}). To see the impact  on the bracket, take two equivalent transformations differing in a reparameterization
\begin{eqnarray}
\tilde\delta_{\zeta}=\delta_{\zeta}+ \delta_{\Delta\zeta}  \, ,  
\end{eqnarray}
where $\delta_{\zeta}$ is a symmetry transformation with a known bracket and $\Delta\zeta=\ell(\zeta,\psi_{i})$ is a sub-leading shift depending on the old parameters $\zeta$ as well as the fields $\psi_i$ of the invariant theory. In our case $\zeta$ will represent the collective parameter $\zeta_n=(\xi^{\mu},\lambda_{\nu},\Lambda^{a}{}_{b},\beta^{\mu\nu})$, containing diffeomorphism, gauge, Lorentz and $\beta$ parameters. The new bracket can be expressed in terms of the old one
 as 
\begin{eqnarray}
\left[\tilde \delta_{\zeta_{1}},\tilde \delta_{\zeta_2}\right]\psi_i=
\left[\delta_{\zeta_1},\delta_{\zeta_2}\right]\psi_i + 
2 \delta_{\zeta_{[1}}\delta_{\Delta\zeta_{2]}}\psi_i +
2 \delta_{\Delta\zeta_{[1}}\delta_{\zeta_{2]}}\psi_i\, .\label{newbracket}
\end{eqnarray}
The first term  gives the old bracket, which is assumed to be known,
\begin{eqnarray}
\left[\delta_{\zeta_1},\delta_{\zeta_2}\right]\psi_i
&=& - \delta_{\zeta_{12}}\psi_i = - \tilde\delta_{\zeta_{12}-\Delta \ell_{12}}\psi_i\, ,
\end{eqnarray}
where $\Delta\ell_{12}=\ell(\zeta_{12},\psi_i)$ is written in terms of the same set of functions $\ell$ introduced above, and
\begin{eqnarray}
\zeta_{12}=g(\zeta_1,\zeta_2,\psi_i) \ ,
\end{eqnarray}
 defines the old bracket for certain known functions $g$.%\footnote{Notice that $\Delta g_{12}^{\mu\nu}=0$ necessarily follows from the requirement that $\beta^{\mu\nu}$ describes rigid transformations and the fact that $\beta_{12}=0$ at leading order.}

The second and third factors in (\ref{newbracket}) can be written as
\begin{eqnarray}
2 \delta_{\zeta_{[1}}\delta_{\Delta\zeta_{2]}}\psi_i +
2 \delta_{\Delta\zeta_{[1}}\delta_{\zeta_{2]}}\psi_i
&=&
\left[\delta_{\zeta_{1}},\delta_{\Delta\zeta_{2}}\right]\psi_i -(1\leftrightarrow 2)\, .
\end{eqnarray}
We can split $\left[\delta_{\zeta_{1}},\delta_{\Delta\zeta_{2}}\right]\psi_i $ into a variation with $\Delta\zeta$ fixed and  $\delta_{\hat\zeta_{12}}\psi_i$ with $\hat\zeta_{12}=\delta_{\zeta_1}(\Delta\zeta_2)$. The former can be cast in terms of the old bracket as
\begin{eqnarray}
\left[\delta_{\zeta_{1}},\delta_{\Delta\zeta_{2}}\right]\psi_i
\left. \vphantom{\frac12}\right|_{\Delta\zeta}  &=& -\delta_{\Delta g_{12}}\psi_i\, ,
\end{eqnarray}
where $\Delta g_{12}= g(\zeta_{1},\Delta\zeta_{2},\psi_i)$, while $\delta_{\hat\zeta_{12}}\psi_i$ is non trivial when  $\Delta\zeta$ is field dependent.
 Taking into account that %$\delta$ and $\tilde\delta$ coincide at zero order and that the parameters $\Delta g_{12}\sim \hat\zeta_{12}\sim{\cal O}(\alpha')$, it holds 
$\delta_{\Delta g_{12}}=\tilde\delta_{\Delta g_{12}}$ and $\delta_{\hat\zeta_{12}}=\tilde\delta_{\hat\zeta_{12}}$ up to ${\cal O}(\alpha'{}^2)$, the bracket after reparameterization turns out to be
\begin{eqnarray}
\zeta^{(new)}_{12}=\zeta^{(old)}_{12} -\Delta \ell_{12} + 2\Delta g_{[12]} - 2 \hat\zeta_{[12]}.\label{newbracket1}
\end{eqnarray}

\section{Bergshoeff-de Roo scheme}\label{bdrs}

The gravitational sector in the DFT scheme is rather unconventional because the metric is not Lorentz invariant. There is however a Lorentz non-covariant redefinition of the vielbein that connects it to a scheme with a Lorentz invariant metric. Actually, inserting in \eqref{fg} the function
\ba
F_{ab}=\frac{a+b}8\left(w_{acd}w_b{}^{cd}+\frac14H_{acd}H_b{}^{cd}\right)+\frac{b-a}8w_{(a}{}^{cd}H_{b)cd}\, , \label{FBdR}
\ea
one obtains the so-called generalized Bergshoeff-de Roo (BdR) scheme \cite{marn}  with this sole redefinition. 
 Since $G_{\mu \nu} = 0$,  the two-form is not redefined with respect to the DFT scheme, and so it maintains a non-trivial Lorentz transformation for any value of $a$ and $b$ (\ref{BosonicLorentz}). This might look surprising, as for the bosonic string (which is obtained with $a=b$), the two-form is normally Lorentz invariant. We will  later clarify how  this is resolved by moving to the Metsaev-Tseytlin scheme.

The bi-parametric $(a\, ,\, b)$ BdR action \cite{marn} generalizes that of \cite{bdr} as follows 
\begin{equation}
    S_{\rm{BdR}}= \int d^D x \sqrt{-g} e^{-2\phi} \left(L^{(0)} + a L^{(1)}_a + b L^{(1)}_b \right) \ ,\label{bdrac}
\end{equation}
where the lowest order Lagrangian $L^{(0)}$ is defined in  (\ref{DSugra}), and the first order one can be written in a flat index notation 
\begin{subequations}
\begin{align}
L^{(1)}_a &= \frac 1 4 H^{a b c} \Omega^{(-)}_{a b c} - \frac 1 8 R^{(-)}_{a b c d} R^{(-) a b c d} \ , \\
L^{(1)}_b &= - \frac 1 4 H^{a b c} \Omega^{(+)}_{a b c} - \frac 1 8 R^{(+)}_{a b c d} R^{(+) a b c d} \ .
\end{align}
\end{subequations}
Defining the torsionful connections $w^{(\pm)}_{a b c} = w_{a b c} \pm \frac 1 2 H_{a b c}$, these expressions contain the Chern-Simons three-forms $\Omega^{(\pm)}_{a b c}$ and Riemann tensors $R^{(\pm)}_{a b c d} $,
\begin{eqnarray}
\Omega^{(\pm)}_{a b c} &=&  w^{(\pm)}_{[\underline{a} d}{}^e D_{\underline{b}}w^{(\pm)}_{\underline{c}] e}{}^d + w^{(\pm)}_{[\underline{a} d}{}^ew^{(\pm)}_{f e}{}^dw_{\underline{b c}]}{}^f + \frac 2 3 w^{(\pm)}_{[\underline{a} d}{}^ew^{(\pm)}_{\underline{b} e}{}^fw^{(\pm)}_{\underline{c}] f}{}^d \ ,\label{omega+-}\\
R^{(\pm)}_{a b c d} &=& 2 D_{[a}w^{(\pm)}_{b]cd} + 2 w_{[a b] }{}^e w^{(\pm)}_{e c d}  + 2 w^{(\pm)}_{[\underline{a} c}{}^e w^{(\pm)}_{\underline{b}]e d} \ .\label{riemann+-}
\end{eqnarray}

Plugging (\ref{FBdR}) into (\ref{tfg}) and using the identities (\ref{identities}-\ref{transftensors}), we find
 \begin{subequations}\label{betat}
\begin{align} 
\delta^{(1)} e_{a b} &= \frac {a + b} 8 \beta_{(a}{}^e \left( w_{b) c d} H_{e}{}^{c d} + H_{b)c d} w_{e}{}^{c d} \right) +  \frac {b - a} 4 \beta_{(a}{}^e \left( w_{b) c d} w_{e}{}^{c d} + \frac 1 4 H_{b)c d} H_{e}{}^{c d} \right) \ , \  \\
\delta^{(1)} b_{a b} &=    (a + b) \left[ \beta^{e c} w_{e [a}{}^d w_{b] c d} - \beta^{e c} w_{[\underline{a} e}{}^d w_{\underline{b}] c d} - \frac 1 2 \beta_{[a}{}^c w_{b] d e} w_c{}^{de} - \frac 1 8 \beta_{[a}{}^c H_{b] d e} H_{c}{}^{d e}\right] \nn\\
& \ \ + \, \frac {b - a} 2 \left[ \beta^{e c} w_{e [a}{}^d H_{b] c d} - \beta^{e c} w_{[\underline{a} e}{}^d H_{\underline{b}] c d} - \frac 1 2 \beta_{[a}{}^c w_{b] d e} H_c{}^{de} - \frac 1 2 \beta_{[a}{}^c H_{b] d e} w_{c}{}^{d e}\right] \ .
\end{align}
\end{subequations}
These expressions are in perfect agreement with our previous results in  equations (3.12) of \cite{bmn}, where they were obtained from the requirement of invariance of the action. Actually, since $\delta^{(1)} L^{(0)}$ is of the form (\ref{d1L0}), namely a sum of terms linear in the equations of motion plus total derivatives, we took $\delta^{(0)} L^{(1)}$  precisely to that form, which allowed us to read the first order transformations (\ref{betat}). We then verified that the total derivatives in $\delta^{(1)} L^{(0)}$ are exactly canceled by those in $\delta^{(0)}L^{(1)}$, implying $\delta^{(0)} L^{(1)} + \delta^{(1)} L^{(0)} = 0$ concluding that the transformations (\ref{betat}) in fact leave the Lagrangian invariant. 
 
We have seen in the previous section that a complete set of symmetries that leave the Lagrangian invariant must necessarily close off-shell. In \cite{bmn} we confirmed this general statement, in particular for the transformations \eqref{betatransf} and \eqref{betat}, which close in combination with the local symmetries of the theory, with the following  $\alpha'$-corrected brackets\footnote{
Here the bracket for $\lambda_{12\mu}$ differs from the one in \cite{bmn} by a trivial transformation proportional to $ \partial_{\mu}\left(\Lambda_{[1}^{ab}\beta_{2]ab}\right)$. }
\begin{subequations}\label{clos}
\begin{align}
\Lambda_{12}{}_{a b} &= 2 \, \beta_{1[a}{}^{c} \beta_{2\, b] c} 
    + 2\, \Lambda_{1 [a}{}^{c} \Lambda_{2\, b] c}
    + 2\, \xi_{[1}^{\mu}\partial_{\mu}\Lambda_{2] a b} \cr
  &\ \ \ -  4\,{F}_{c [a} \beta_{[1}{}_{b]}{}_{d} \beta_{2]}^{c d}  
- 4\, {F}^{c d} {{\beta_{1}}}_{c [a} {{\beta_{2}}}_{b] d}  
- \left[\frac{a+b}{4} \,  {H}_{ec d}   
+\frac{b-a}{2} \, {w}_{ec d}  \right]\beta^e_{[1  [a}\, {D}_{b]} \Lambda_{2]}^{c d} 
\cr
& \ \ \ - \left[(a+b)\, {w}^{e f}{}_{d}  + \frac{b-a}{2} {H}^{e f}{}_{d}
\right] \left({w}_{[a}{}^{c d}  + {w}^{cd}{}_{[a} \right) \beta_{[1 b] e} \beta_{2] c f} \ ,  \\
\lambda_{12 \mu}&=
4 \,\xi_{[1}^{\nu} \partial_{[\nu}\lambda_{2] \mu]}  
- \, \frac{a+b}{4} \left(
 \partial_{\mu}\Lambda_{[1}^{a b}\;  \beta_{2]}{}_{a b} - \Lambda_{[1}^{a b}\; \partial_{\mu} \beta_{2]}{}_{a b} 
 \right)  \cr
& \ \ \ - (b-a) \left( \Lambda_{[1}^{a b} \partial_{\mu}\Lambda_{2]}{}_{a b} 
+ \beta_{[1}^{a b} \partial_{\mu}\beta_{2]}{}_{a b} 
\right) \ , \\
\xi_{12}^{\mu} &= 2 \xi^{\nu}_{[1} \partial_{\nu}\xi_{2]}^{\mu} + \beta_{[1}^{\mu\nu} \lambda_{2] \nu} \; , \label{brackets}
\end{align}
\end{subequations}
where $\xi^\mu$ stands for the vector parameter of diffeomorphisms and $\lambda_\mu$ for the one-form gauge parameter of $b_{\mu\nu}$. The presence of diffeomorphisms in the closure of the algebra might appear in contradiction with the results in Section \ref{InvarianceClosure}, where we found that a complete set of symmetries that leave a Lagrangian invariant must close among each other. Indeed, the Lagrangian is not invariant but transforms as a scalar with respect to the Lie derivative. However, this apparent contradiction is easily resolved by noting that only the subset of trivial diffeomorphisms $\xi^\mu = \beta^{\mu \nu} X_\nu$, i.e. those that leave the Lagrangian invariant, are required to close the algebra. 

We end this section by noting that the $\beta$ transformations \eqref{betat}
admit first order deformations, as discussed in subsection \ref{4param}. For any non vanishing value of the set of parameters ($\gamma, \chi, \rho, \kappa$), the Lagrangian would not to be invariant, but it would transform up to a total derivative, which then implies that the action is invariant. Activating these deformations would then require using the equations of motion to close the symmetry algebra, as discussed in subsection \ref{InvarianceClosure}.

\section{Metsaev-Tseytlin scheme }

The functions $F_{ab}$ and $G_{\mu\nu}$ that connect the DFT with the Metsaev-Tseytlin scheme (MT) are 
\begin{subequations} \label{redefMT}
\begin{align}
F_{ab}\ &= \ \frac{a+b}8\left(w_{acd}w_b{}^{cd}+\frac34H_{acd}H_b{}^{cd}\right)+\frac{b-a}8w_{(a}{}^{cd}H_{b)cd} \;, \\
G_{\mu\nu} \ &= - \ \frac{a+b}{4} \left( {\cal D}^{\rho} H_{\rho \mu\nu} + H_{[\mu}{}^{c d} w_{\nu] c d}\right)\, ,
\end{align}
\end{subequations}
where ${\cal D}_\mu$ is the curved version of (\ref{calD}). These redefinitions take the bi-parametric action in the DFT scheme to the following form \cite{marn}
\begin{equation}
    S_{\rm{MT}}=  \int d^D x \sqrt{-g} e^{-2\phi}\left( L^{(0)} +  L^{(1)}_-+  L^{(1)}_+ + {\cal D}_\mu V^\mu  \right)\ ,    \label{mtac}
\end{equation}
where  $L^{(0)}$ is the universal two-derivative Lagrangian (\ref{DSugra}), and the first order corrections induced from DFT contain the standard MT first order terms \cite{MT} (see Appendix \ref{AppA} for definitions)
\begin{eqnarray}
L^{(1)}_-&=&\frac {a-b}4 H^{\mu\nu\rho}\Omega_{\mu\nu\rho}\, ,\label{min}\\
L^{(1)}_+&=&-\frac{a+b}8\left[R_{\mu\nu\rho\sigma}R^{\mu\nu\rho\sigma}-\frac12H^{\mu\nu\rho}H_{\mu\sigma\lambda}R_{\nu\rho}{}^{\sigma\lambda}+\frac1{24}H^4-\frac18H_{\mu\nu}^2H^{2\mu\nu}\right]\, ,\label{oep}
\end{eqnarray}
and includes in addition a total derivative ${\cal D}_\mu V^\mu$ with
\be
V^\mu =  \frac{a+b}{8} H^{\mu \nu \rho}{\cal D}^{\sigma}H_{\sigma \nu \rho} \ . \label{mttd}
\ee
Of course, since this is what descends directly from DFT, we expect the Lagrangian to be $\beta$ invariant only with the inclusion of this term. 

The terms proportional to $a+b$, which are common to the bosonic and heterotic strings, contain even powers of $H_{\mu \nu \rho}$ and are then invariant with respect to an exchange of sign of the two-form, $b_{\mu \nu} \to - b_{\mu \nu}$. In addition, they are purely metric, in the sense that the vielbein enters into these terms only through the metric. Notice also that in this case, the two-form is Lorentz invariant, as expected for the bosonic string in this scheme.

On the other hand, the term proportional to $a-b$ involves the Chern-Simons form contained in the heterotic string, which is linear in $H_{\mu \nu \rho}$, and hence is odd with respect to the exchange of sign of the two-form. This is a frame-like contribution, as it cannot be written purely in terms of the metric. Lorentz invariance of $L_-^{(1)}$ requires the non-standard Green-Schwarz Lorentz transformation of the two-form \cite{marn}
\ba
\delta_\Lambda b_{\mu\nu}=-\frac {a-b} 2\partial_{[\mu}\Lambda_c{}^dw_{\nu]d}{}^c\, .
\ea

$\beta$ transformations in the MT scheme follow from inserting (\ref{redefMT}) into  \eqref{tfg}
\begin{eqnarray}
\delta^{(1)}_\beta e_{ab} = \tau_{(a b)} + f_{[a b]}\, ,\qquad\qquad
\delta^{(1)}_\beta b_{a b} =  \sigma_{a b} + \nabla_{[a}t_{b]}\, ,\qquad\qquad
\delta^{(1)}_\beta \phi =  \frac12 \tau_{a}{}^{a} \;,\label{transfmt}
\end{eqnarray}
with
\begin{subequations}
\begin{align}
\tau_{a b}
 =& - \frac{a+b}{2} \left(
 \nabla{}_{d}\beta_{c(a} H_{b)}{}^{c d} + \frac12    \beta_{c(a} {\cal D}{}_{d}{H_{b)}{}^{c d} }
  \right) - \frac{b-a}{4}   \beta_{c(a}  \left({w}_{b) d e} {w}^{c d e} + \frac14 H_{b) d e} H{}^{c d e}  \right)  \,,
\\
 f_{a b} =& \ - \frac{a+b}{8}\beta_{c [a} \left( \vphantom{\frac12}  
 {w}_{b] d e}  H^{c d e} -  H_{b] d e}  {w}^{c d e} + 2  {\cal D}^{d}{H_{b] d}{}^{c}} \right) \, ,
 \label{fab}\\
 \sigma_{a b} =& \
(a+b) \left[ -\frac12 \beta{}^{c d} \left( R_{a b c d}  + \frac12 H_{a c e} H_{b d}{}^{e}\right)
+ \beta_{c [a} \left(R_{b]}{}^{c} + 2 \nabla_{b]}\nabla^{c}\phi +  \frac{1}{4} H_{b] d e} H{}^{c d e} \right)    \right]\nn\\&
+ \frac{b - a}{2} \left[ \beta_{c d} H{}^{d e}{}_{ [a } \left(  {w}_{b]}{}^{c}{}_{e} -  {w}{}^{c}{}_{ b] e}  \right)
+\frac{1}{2} \beta_{c [a}\left( {w}_{b] d e}  H{}^{c d e} +  H_{b] d e}{w}{}^{c d e}\right)\right] \;,\\
t_{a} =& \ (a+b)\; \beta{}^{b c} w_{a b c} \label{ta} \;.
\end{align}
\end{subequations}
Written in this form, equations \eqref{transfmt} highlight the observation made in subsection \ref{BracketRedef}, about redefinitions of the symmetry parameters. In this case, we see that the $f_{[ab]}$ piece of the $\beta$ transformations can be absorbed into redefinitions of the Lorentz  parameter and $t_a$ into (the flat version of) the one-form that parameterizes the gauge transformations of the Kalb-Ramond field. We will assume this redefinition of the parameters from now on. In particular, this means that we will have to take care of the redefined brackets discussed in subsection \ref{BracketRedef} when verifying closure of the symmetry algebra.

Since the terms in ${L}^{(1)}_+$ are purely metric, we expect the corresponding $\beta$ transformations to be purely metric as well. Indeed, performing the parameter redefinitions mentioned in the previous paragraph (i.e. ignoring $f_{[a b]}$ and $t_a$), we obtain the following $\beta$ transformations for the metric and the two-form  
\begin{eqnarray}
\delta^{(1)}_\beta g_{\mu\nu}\!\!\!\! &=& \!\!\!\! (a+b)  \left(
 \nabla{}^{\sigma}\beta{}_{(\mu}{}^{\lambda} H{}_{\nu) \lambda\sigma} + \frac12    \beta{}_{(\mu}{}^{\lambda}  {\cal D}{}^{\sigma}{H{}_{\nu) \lambda\sigma} }
  \right)
+ \frac{a-b}{2}   \beta{}^{\lambda}{}_{(\mu}   \left({w}_{\nu) c d} {w}_{\lambda}{}^{c d} + \frac14 H{}_{\nu) \rho \sigma} H{}_{\lambda}{}{}^{\rho\sigma}  \right)\, , \nn \\
\delta^{(1)}_{\beta} b_{\mu\nu} \!\!\!\!&=&\!\!\!\!  \
(a+b) \left[ -\frac12 \beta{}^{\rho \sigma} \left( R_{\mu \nu \rho \sigma}  + \frac12 H_{\mu \rho \tau} H_{\nu \sigma}{}^{\tau}\right)
+ \beta_{\rho [\mu} \left(R_{\nu]}{}^{\rho} + 2 \nabla_{\nu]}\nabla^{\rho}\phi +  \frac{1}{4} H_{\nu] \sigma \tau} H{}^{\rho \sigma \tau} \right)    \right]\nonumber\\&&\!\!\!\!
- \frac{a - b}{2} \left[ \beta_{c d} H{}^{d e}{}_{ [\mu } \left(  {w}_{\nu]}{}^{c}{}_{e} -  {w}{}^{c}{}_{ \nu] e}  \right)
+\frac{1}{2} \beta_{c [\mu}\left( {w}_{\nu] d e}  H{}^{c d e} +  H_{\nu] d e}{w}{}^{c d e}\right)\right] \ .
\end{eqnarray}
As expected, the terms proportional to $a+b$ can be cast in purely metric expressions, with spacetime greek indices. Instead, those proportional to $a-b$, involve spin connections that necessarily require the vielbein,  as they cannot be expressed only in terms of the metric.

By construction, since we derived them from DFT, these $\beta$ transformations leave the Lagrangian invariant. Hence, they close off-shell, as we proved in subsection \ref{InvarianceClosure}. On the other hand, as discussed in  \ref{4param}, if one is interested in invariance of the action, instead of focusing on invariance of the Lagrangian, then these transformations can be supplemented with the four-parameter family of deformations (\ref{deformparam2}), that leave the action invariant and determine that the symmetry algebra closes on-shell.

After reparameterization, the brackets in this scheme  can be computed from the corresponding ones in the BdR scheme, following the discussion in subsection \ref{BracketRedef}. Details of the computation can be found in Appendix \ref{AppB}. The brackets $\xi^{\mu}_{12}$ and $\beta^{\mu\nu}_{12}$ are not modified, while those for  gauge and Lorentz transformations turn into
\begin{eqnarray}
\lambda_{\mu}^{12}  \!\!\!
&=&  \!\!\!
- 4 \,\xi_{{}_{[1}}{}^{\nu} \partial_{[\mu}\lambda_{{}_{2]}}{}_{\nu]} 
- (b-a) \left( \Lambda_{[1}^{a b} \partial_{\mu}\Lambda_{2]}{}_{a b} 
+ \beta_{[1}^{a b} \partial_{\mu}\beta_{2]}{}_{a b} 
\right) - \, \frac{a+b}{4} \left(
\partial_{\mu} \Lambda_{[1}^{a b}\;  \beta_{2] a b} 
- \;  \Lambda_{[1}^{a b}\;  \partial_{\mu} \beta_{2]a b} 
\right) \nn\\
\Lambda^{12}_{ab} \!\!\!
&=& \!\!\!
\Lambda^{(0)12}_{ab} 
- 4\,{K}_{c [a} \beta^{[1}{}_{b]}{}_{d} \beta^{2]}{}^{c d}  
- 4\, {K}^{c d} \beta^{[1}_{c [a} {{\beta^{2]}}}_{b] d} - 4 \Lambda^{[1}_{[a}{}^{c} f^{2]}_{b]c} - \frac{b-a}{2} \, {\omega}^{e}{}_{c d} {{\beta^{[1}}}_{e [a}\, {D}_{b]}{\Lambda^{2]c d}}\nn\\
&& \!\!\!-
\left[ 2(a+b) \left( {\omega}^{e f}{}_{d}  \omega^{cd}{}_{[a}
+\frac14 {H}^{e f}{}_{d}  H{}^{cd}{}_{[a} \right)
 + \frac{b-a}{2}  {H}^{e f}{}_{d}
\left({\omega}_{[a}{}^{c d}  + {\omega}^{cd}{}_{[a} \right)
 \right] \beta^{[1}_{b] e} \beta^{2]}_{c f}\; \;
,\label{bracketMT}
\end{eqnarray}
where 
\begin{eqnarray}
K_{ab}=\frac{a+b}{4}\left(\vphantom{\frac12}
R_{ab} + 2 \nabla_{a}\nabla_{b}\phi + \frac14 H_{acd} H_{b}{}^{cd}
\right)+\frac{b-a}8\omega_{(a}{}^{cd}H_{b)cd}\ .
\end{eqnarray}
Despite having computed the brackets systematically from the one obtained in \cite{bmn}, we have verified explicitly that they lead exactly to off-shell closure of the symmetry algebra.

\subsection{Invariance of the Metsaev-Tseytlin Lagrangian}\label{MTsection}
Here we present an explicit derivation of the invariance of the bi-parametric MT Lagrangian (\ref{mtac})-(\ref{mttd}) with respect to the transformations (\ref{transfmt})-(\ref{ta}). We will first focus on the terms proportional to $a+b$ (\ref{oep}), and then we deal with those proportional to $a-b$ (\ref{min}).
\subsubsection{Invariance of the even parity sector }\label{subl1+}

 The deformations of the $\beta$ transformations that preserve the even sector of the MT action were systematically derived  from DFT, namely
\begin{subequations} \label{TransfPlus}
\begin{align}
\delta^{(1)}_\beta e_{ab} =& - \frac{a+b}{2} \left(
 \nabla{}_{d}\beta_{c(a} H_{b)}{}^{c d} + \frac12    \beta_{c(a} {\cal D}{}_{d}{H_{b)}{}^{c d} }
  \right) + \gamma \beta_{c(a} {\cal B}_{b)}{}^{ c} +  \chi g_{a b} \beta_{c d} {\cal B}^{ c d}  \\
\delta^{(1)}_\beta b_{a b} =&  (a+b) \left[ -\frac12 \beta{}^{c d} \left( R_{a b c d}  + \frac12 H_{a c e} H_{b d}{}^{e}\right)
+ \beta_{c [a} \left(R_{b]}{}^{c} + 2 \nabla_{b]}\nabla^{c}\phi +  \frac{1}{4} H_{b] d e} H{}^{c d e} \right)    \right] \nonumber \\
& - \gamma \beta^c{}_{[a} {\cal E}_{b]c}  -  \chi \beta_{a b} {\cal E}_c{}^c  \ .
\end{align}
\end{subequations}
We consider here only the deformations proportional to the parameters $\gamma$ and $\chi$, as they preserve the expected parity of the transformations. Since this sector is common to the bosonic and heterotic strings, the transformations can be cast in a purely metric formulation. Indeed, as expected, we find
\begin{subequations}
\begin{align}
\delta^{(1)}_\beta g_{\mu\nu} =& - (a+b)  \left(
 \nabla{}^{\sigma}\beta{}^{\lambda}{}_{(\mu} H{}_{\nu) \lambda\sigma} + \frac12    \beta^{\lambda}{}_{(\mu}  {\cal D}{}^{\sigma}{H{}_{\nu) \lambda\sigma} }
  \right) + 2 \gamma \beta_{\rho(\mu} {\cal B}_{\nu)}{}^{ \rho} + 2 \chi g_{\mu \nu} \beta_{\rho \sigma} {\cal B}^{\rho\sigma}\, , \label{delta1mt}\\
\delta^{(1)}_{\beta} b_{\mu\nu} =& 
(a+b) \left[ -\frac12 \beta{}^{\rho \sigma} \left( R_{\mu \nu \rho \sigma}  + \frac12 H_{\mu \rho \tau} H_{\nu \sigma}{}^{\tau}\right)
+ \beta_{\rho [\mu} \left(R_{\nu]}{}^{\rho} + 2 \nabla_{\nu]}\nabla^{\rho}\phi +  \frac{1}{4} H_{\nu] \sigma \tau} H{}^{\rho \sigma \tau} \right)    \right]\nn\\&
-  \gamma \beta^\rho{}_{[\mu} {\cal E}_{\nu]\rho}  -  \chi \beta_{\mu \nu} {\cal E}_\rho{}^\rho  \ . \label{delta2mt}
\end{align}
\end{subequations}
By construction, for $\gamma = \chi = 0$ these transformations must leave the even parity Lagrangian  invariant. Here we want to show this explicitly, and for this we must verify that the following equation holds
\be
\delta_\beta^{(1)} L^{(0)} + \delta^{(0)}_\beta (L_+^{(1)} + {\cal D}_a V^a) = 0\, .\label{var1}
\ee

To this end, we first replace  (\ref{d1L0}) in the first term and the fact that $[\delta^{(0)}_\beta \, , \, {\cal D}_a] W^a = 0$ for any $W^a$, to re-write (\ref{var1}) as
\be
\delta^{(0)}_\beta L^{(1)}_+ = \underbrace{- {\cal E}{}^{ab} \; \delta^{(1)}_\beta  e_{ab} - {\cal B}^{ab} \; \delta^{(1)}_\beta b_{ab}  \vphantom{\left[\frac 1 2\right] } }_{(A)}
\underbrace{- {\cal D}_{a}\left[2\; {\cal D}_{b}\delta^{(1)}_\beta e^{ab} -\frac12 H^{abc} \delta^{(1)}_\beta b_{bc} + \delta^{(0)}_\beta V^a\right]}_{(B)} \ . \label{var0}
\ee
Next, we  explicitly compute the lowest order variation of $L^{(1)}_+$ (\ref{oep}) in the LHS.  For this it is useful to introduce the following identities
\begin{eqnarray}
\delta^{(0)}_{\beta} \left( R_{abcd} R^{abcd} \right)  &=&  \left[\vphantom{\frac12}-4 \nabla_{a}\left(\beta_{b}{}^{e} H_{e c d} \right) +\left(\nabla^{e}\beta_{ab} - 2 \nabla_{a}\beta_{b}{}^{e} \right)H_{e cd} \right]R^{abcd}\, , \ \ \ \ \ \ \ \ \ \ \ \  \\
 \delta^{(0)}_\beta\left( -\frac12 H_{eab} H^{e}{}_{cd} R^{abcd}\right)  &=&
  \; H_{e}{}^{ab} H^{ecd} \left[ \nabla_{a}\left(\beta_{b}{}^{f}H_{f cd}\right)+\left(\nabla_{a} \beta_{c}{}^{f} - \frac12 \nabla^{f}\beta_{ac}\right)H_{bdf}\right]\, \nn
  \\
&&  +  \left[\vphantom{\frac12}\nabla^{e}\beta_{ab}+ 2 \nabla_{a}\beta_{b}{}^{e} \right]H_{ecd} R^{abcd}\, , \\
\delta^{(0)}_{\beta} \left( - \frac18 H_{a e f} H_{b}{}^{e f} H^{a}{}_{c d} H^{b c d}\right)  &=&  \frac12 \left[\vphantom{\frac12} \nabla_{a}\beta_{cd} +2 \nabla_{c}\beta_{da} \right]H_{b}{}^{cd}  H^{a ef} H^{b}{}_{ef}\, ,\\
\delta^{(0)}_{\beta} \left( \frac{1}{24} H_{abc} H^{ade} H^{b}{}_{e f} H^{cf}{}_{d}  \right)  &=& -\frac12 \nabla_{a}\beta_{bc}  H^{ade} H^{b}{}_{e f} H^{cf}{}_{d} \, ,
\end{eqnarray}
which readily lead to the following result 
\ba
\delta^{(0)}_\beta L_+^{(1)} &=& \frac{a+b}{2} R^{abcd} \left[\nabla_{a}\left(\beta_{b}{}^{e} H_{e c d}\right) - \frac12 \nabla_{e}\beta_{ab} H^{e}{}_{cd}\right] \nn\\
&& - \frac{a+b}8 \;H_{f}{}^{ab} H^{fcd} \left[\nabla_{a}\left(\beta_{b}{}^{e} H_{e c d}\right) + \frac32 \nabla_{[e}\beta_{ab]} H^{e}{}_{cd}\right] \, .\label{d0l+1}
\ea
Following the steps detailed in Appendix \ref{AppC}, this can be rewritten as 
\be
\delta^{(0)}_\beta L_+^{(1)} = (a+b) \left[(A') + (B')\right] \ ,\label{VarAppendix}
\ee
with
\begin{align} 
(A')&=  - {\cal E}_{ab}  \nabla_{f}\beta^{ae} H_{e}{}^{bf}  
- {\cal B}_{ab}   \left(R^{abef} \beta_{ef}   - \beta^{da} H^{2b}{}_{d} +  \frac12 \beta_{ef} H^{bf}{}_{d}H^{aed}  \right)
-2 \beta^{d}{}_{a} {\cal B}^{ab} {\cal E}_{db}  \ ,
 \\
(B') &=  - {\cal D}_{a}\left[\vphantom{\frac12} \beta_{b e} \left( H_{cd}{}^{e} R^{abcd}  - \frac12 H_{cd}{}^{a} \left(R^{be cd} -\frac12H^{cef}H_f{}^{db} \right) + 2 \left(\vphantom{e^{\frac12}} R^b{}_{c} 
+ 2 \nabla^{b}\nabla_{c}\phi \right) H^{eca} \right)
\right] \; , \nn
\end{align}
which is  closer in structure to (\ref{var0}).  It is now easy to recover the conditions for invariance of the Lagrangian and the action by comparing (\ref{VarAppendix}) with (\ref{var0}). 
\begin{itemize}
\item For the action to be invariant, it is enough to verify 
\be (A) = (a+b) (A') \label{invaction} \ .\ee 
By definition, when inserting   (\ref{TransfPlus}) into $(A)$, the terms depending on $\gamma$ and $\chi$ simply drop out. The rest of the terms can be rapidly shown to verify this identity. As a consequence, the action is invariant for whatever choice of the parameters  $\gamma$ and $\chi$.
\item For the Lagrangian to be invariant, on top of (\ref{invaction}), also
\be (B) = (a+b) (B') \label{invLag} \ ,\ee
must hold. This is the imposition that the total derivatives that emerge from the variation of the lowest order Lagrangian are exactly canceled by those arising in the variation of the first order Lagrangian. In this case, when (\ref{TransfPlus}) are inserted in $(B)$, the identity (\ref{invLag}) can be shown to hold for the specific choice $\gamma = \chi = 0$ (see Appendix \ref{AppC}).
\end{itemize}

In summary, we have shown that the transformations (\ref{TransfPlus}) leave the even parity sector of the MT action invariant for any choice of the parameters $\gamma$ and $\chi$. Instead, the Lagrangian was shown to be invariant for the specific choice $\gamma = \chi= 0$, which is the case that descends directly from DFT.

\subsubsection{Invariance of the odd parity sector }\label{subl1-}

We now repeat the analysis for the terms proportional to $a-b$, namely $L^{(1)}_{-}$. We have seen that, up to parameter redefinitions, the first order transformations that leave the odd parity piece of the action invariant, and respect the parity of the fields accordingly, are
\begin{subequations}\label{VarMenos}
\begin{align}
\delta^{(1)}_\beta e_{ab} =&  \frac{a - b}{4}   \beta_{c(a}  \left({w}^{2\ c}_{b)}  + \frac14 H^{2\ c}_{b)}   \right) + \kappa  \beta_{(a}{}^c {\cal E}_{b)c} \, ,\\
\delta^{(1)}_\beta b_{a b} =&  - \frac{a - b}{2} \left[ \beta_{c d} H{}^{d e}{}_{ [a } \left(  {w}_{b]}{}^{c}{}_{e} -  {w}{}^{c}{}_{ b] e}  \right)
+\frac{1}{2} \beta_{c [a}\left( ({wH})_{b] } {}^{c } +  (Hw)_{b] }{}^{c}\right)\right] + \rho \beta_{c[a} {\cal B}^{c}{}_{b]} \, .
\end{align}
\end{subequations}
To verify explicitly the invariance of the Lagrangian, we must check that the following equation holds
\be
\delta_\beta^{(1)} L^{(0)} + \delta^{(0)}_\beta L_-^{(1)}  = 0\, .
\ee
Replacing  (\ref{d1L0}) in the first term leads to 
\be
\delta^{(0)}_\beta L^{(1)}_- = \underbrace{- {\cal E}{}^{ab} \; \delta^{(1)}_\beta  e_{ab} - {\cal B}^{ab} \; \delta^{(1)}_\beta b_{ab}    \vphantom{\left(\frac 1 2\right)}    }_{(C)} 
\underbrace{- {\cal D}_{a}\left[2\; {\cal D}_{b}\delta^{(1)}_\beta e^{ab} -\frac12 H^{abc} \delta^{(1)}_\beta b_{bc}\right]}_{(D)} \ . \label{var0Menos}
\ee
The explicit variation of $L^{(1)}_-$ in the LHS is shown  in Appendix \ref{AppC} to take the form
\be
\delta_\beta^{(0)} L^{(1)}_{-} = (a-b) \left[(C') +  (D')\right] \ , \label{d0LCS}
 \ee
with
\begin{eqnarray}
(C')&=&-
\frac12 {\cal E}{}^{ab} \beta_{ca}   \left(w^2{}_{b}{}^{c}+\frac14 H^2{}_{b}{}^{c} \right) -  \frac12 {\cal B}{}^{ab} \left[ \beta_{cd} H^{de}{}_{a} \left(w_{b}{}^{c}{}_{e} -w^{c}{}_{be}\right)+ \frac12 \beta_{ca} (Hw + wH)_{b}{}^{c}\right]\, ,  \nn \\
(D')&=& \frac12  {\cal D}_{b}\left[\vphantom{{\frac12}^{\frac12}} \right.
H^{abc} \left(H_{aef} w_{[dc]}{}^{e}\beta^{fd}
 -\frac14  \beta_{cd}  (Hw+wH)_{a}{}^{d} \right)
+  \beta_{ca}w^{ab}{}_{d}  \left(\frac14 H^{2}{}^{cd} - w^{2}{}^{cd}\right)
 \left.\vphantom{{\frac12}^{\frac12}} \right] \, , \nn
\end{eqnarray}
 which leaves us very close to our purpose. It is now easy to recover the conditions for invariance of the Lagrangian and the action by comparing (\ref{d0LCS}) with (\ref{var0Menos}). 
\begin{itemize}
\item For the action to be invariant, it is enough to verify 
\be (C) = (a-b) (C') \label{invaction2} \ .\ee 
By definition, when inserting   (\ref{VarMenos}) into $(C)$, the terms depending on $\kappa$ and $\rho$ simply drop out. The rest of the terms can be rapidly shown to verify this identity. As a consequence, the action is invariant for whatever choice of the parameters  $\kappa$ and $\rho$.
\item For the Lagrangian to be invariant, on top of (\ref{invaction2}), also
\be (D) = (a-b) (D') \label{invLag2} \ ,\ee
must hold. This is the imposition that the total derivatives that emerge from the variation of the lowest order Lagrangian are exactly canceled by those arising in the variation of the first order Lagrangian. In this case, when (\ref{VarMenos}) are inserted in $(D)$, the identity (\ref{invLag2}) can be shown to hold for the specific choice $\kappa = \rho = 0$ (see Appendix \ref{AppC}).
\end{itemize}

In summary, we have shown that the transformations (\ref{VarMenos}) leave the odd parity sector of the MT action invariant for any choice of the parameters $\kappa$ and $\rho$, while the Lagrangian is left invariant for the specific choice $\kappa = \rho= 0$, which is the case that descends directly from DFT.

\section{Conclusions}
In this paper we derived the $\beta$ symmetry transformation rules of the supergravity fields to first order in $\alpha'$ from the frame-like formulation of DFT. The fact that DFT implies $\beta$ symmetry, in turn means that this symmetry is a necessary condition for DFT uplifts of supergravity interactions. It is not sufficient because $\beta$ symmetry in supergravity requires isometries, whereas in DFT it is realized through the introduction of dual derivatives. The road from supergravity to DFT would require relaxing the assumption of isometries, namely the condition $\beta^{\mu \nu} \partial_\nu \dots = 0$, and extending supergravity with interactions containing dual derivatives, to compensate the failure. This observation might become important when assessing the possibility of finding a DFT formulation for the quartic Riemann terms in maximal supergravity, which currently faces an obstruction, at least in the standard frame-like formulation \cite{hw}. Actually, the duality structure of the  ${\cal O}(\alpha'^3)$ interactions of type II strings \cite{garou} is not well understood yet.   Recent progress in this direction was achieved in \cite{wulff}, where these couplings were fixed at fifth order in fields using $O(d,d)$ symmetry.
 It would then be interesting to study the $\beta$ symmetry of these higher-derivative terms. 

Having derived the $\beta$ transformations systematically from DFT, we then moved to more conventional supergravity schemes. We first showed that the field redefinitions that connect DFT with the generalized Bergshoeff-de Roo scheme of supergravity, allowed us to recover and confirm all the results in \cite{bmn}. Additionally, we performed further redefinitions to reach the Metsaev-Tseytlin scheme. For completion, we showed through explicit computations in this scheme that the MT Lagrangian is exactly $\beta$ invariant for the bosonic and heterotic strings to order $\alpha'$, and that the transformations close off-shell, as expected.

\bigskip

{\bf Acknowledgements:} We warmly thank C. Hull and D. Waldram for hospitality and discussions at Imperial College, where part of this project was developed. We also thank O. Hohm for pointing out reference \cite{hohmbonezzi}. Support by Consejo Nacional de Investigaciones Cient\'ificas y T\'ecnicas (CONICET), Agencia Nacional de Promoci\'on Científica y T\'ecnica (ANPCyT), Universidad de La Plata (UNLP) and Universidad de Buenos Aires (UBA) is also gratefully acknowledged. 

\bigskip

{\bf Note 1:} Our results confirm those of our previous paper \cite{bmn} and are in direct contradiction with supposed obstructions pointed out in  \cite{garousi}-v5  and \cite{garousi2}-v2. Contrary to what is stated in \cite{garousi2}-v2, we confirm here, through explicit computations in subsection \ref{MTsection}, that the  ${\cal O}(\alpha')$ bulk Lagrangian of the heterotic string is exactly invariant under appropriate deformations of the $\beta$ transformations.  We also disagree with statements in  \cite{garousi}-v5, where it is asserted that there is a unique $\beta$ transformation at ${\cal O}(\alpha’)$ that leaves the bulk bosonic action invariant and that it does not form a closed symmetry algebra. The transformation found in equation (31) of \cite{garousi}-v5, corresponds to the particular choice $a=b=-\alpha’$, $\gamma =-\frac{\alpha’}{2}$ and $\chi=0$ in our equations (\ref{delta1mt})-(\ref{delta2mt}). Actually, $\beta$ transformations at ${\cal O}(\alpha’)$  turn out not to be unique, but there is a two-parameter family of deformations in the bosonic string that lead to on-shell closure of the symmetry algebra. 

{\bf Note 2:} After our manuscript appeared in arXiv:2307.02537 [hep-th], the author of \cite{garousi2} found an error  in his calculations, and  corrected the results, producing the revised outcome \cite{garousi3}.

\begin{appendix}
\section{Notation and definitions}\label{AppA}
We use  $\mu,\nu,\rho,\dots$ and  $a,b,c,\dots$ indices for space-time and tangent space coordinates, respectively. The infinitesimal Lorentz transformation of the vielbein is
\be
\delta_\Lambda e_\mu{}^a = e_\mu{}^b \Lambda_b{}^a \ \, .
\ee
The  spin connection 
\be
w_{c a b} = e^\mu{}_c\left(\partial_\mu e_{\nu a}e^\nu{}_b -\Gamma^\rho_{\mu\nu} e_{\rho a} e^\nu{}_b\right)\, ,\qquad {\rm with }\ \ \ \ \  \Gamma^\rho_{\mu\nu}=\frac12 g^{\rho\sigma}\left(\partial_\mu g_{\sigma\nu}+\partial_\nu g_{\mu\sigma}-\partial_\sigma g_{\mu\nu}\right)\, ,
\ee
transforms as
\be
\delta_\Lambda w_{c a b} =  D_c \Lambda_{a b} + w_{d a b}  \Lambda^d{}_c  + 2 w_{c d [b} \Lambda^d{}_{a]}\, ,
\ee
and hence, it turns flat derivatives $D_a$ into covariant flat derivatives $\nabla_a$ as
\be
\nabla_a T_b{}^c = D_a T_b{}^c+ w_{a b}{}^d T_d{}^c-w_{ad}{}^cT_b{}^d \ , \ \ \ D_a = e^\mu{}_a \partial_\mu\, .
\ee 
The Christoffel connection $\Gamma^\rho_{\mu\nu}$ turns spacetime  partial into covariant derivatives as
\be
\nabla_\mu T_{\rho}{}^\sigma
=\partial_\mu T_{\rho}{}^\sigma-\Gamma_{\mu\rho}^\lambda T_\lambda{}^\sigma +\Gamma_{\mu\lambda}^\sigma T_\rho{}^\lambda\, .
\ee

The Riemann tensor 
\be
R^\mu{}_{\nu\rho\sigma}=\partial_\rho\Gamma^\mu_{\nu\sigma}-\partial_\sigma\Gamma^\mu_{\nu\rho}+\Gamma^\mu_{\rho\lambda}\Gamma^\lambda_{\nu\sigma}-\Gamma^\mu_{\sigma\lambda}\Gamma^\lambda_{\nu\rho}\, ,
\ee
with flat space indices is defined as
\be
R_{a b c d} = 2 D_{[a}w_{ b] c d} + 2 w_{[a b]}{}^e w_{e c d} + 2 w_{[\underline{a} c}{}^e w_{\underline{b}] e d}\, .
\ee
While the symmetry $R_{a b c d} = R_{[ab] [cd]}$ is manifest, other symmetries of the Riemann tensor are hidden and  determine the Bianchi identities
\be
R_{a b c d} = R_{c d a b} \ , \ \ \  \ R_{[a b c] d} = 0 \\ , \ \ \  \ \nabla_{[a}R_{b c] d e} = 0  . \label{BI2}
\ee
The Ricci tensor and scalar curvature are given by the traces
\be
R_{a b} = R^c{}_{a c b} \ , \ \ \  R = R_a{}^a \ .  \ \ \ 
R_{[a b]} = 0\ . \label{BI3}
\ee
In terms of the torsionful spin connection 
\be
w^{(\pm)}_{abc}=w_{abc}\pm\frac12 H_{abc}\, ,
\ee
the Riemann tensor defined in \eqref{riemann+-} satisfies the Bianchi identities 
\be
  R^{(\pm)}_{abcd}\ =\  R^{(\mp)}_{cdab} \, , \qquad
  R^{(\pm)}_{[abc]d}\ =\  \pm\frac13 \nabla_{d} H_{abc} +\frac12 H_{[ab}{}^{e} H_{c]de}\, .
\ee
The curvature of the 2-form 
\be
H_{a b c} = 3\, e^\mu{}_a e^\nu{}_b e^\rho{}_c\, \partial_{[\mu} b_{\nu\rho]} \ ,  \ \ \ \ 
 \ee
obeys the Bianchi identity
\be
\nabla_{[a} H_{bcd]} = 0 \ . \label{BI1}
\ee
We define contractions of powers of the 3-form and spin connection as follows:
\ba
H^2&=& H_{abc}H^{abc}\, ,\qquad \qquad \ H^2_{ab}=H_{acd}H_b{}^{cd}\, ,\qquad\ \ \    H^4=H^{abc}H_{ad}{}^e H_{be}{}^{f}H_{cf}{}^d\, ,\\
w^2_{ab}&=&w_{acd}w_b{}^{cd}\, ,\qquad \ \ \ (wH)_{ab}=w_{acd}H_b{}^{cd}\, ,\qquad \ \ \ (Hw)_{ab}=H_{acd}w_b{}^{cd} \ .
\ea
The Chern-Simons form 
\be
\Omega_{\mu\nu\rho}= w_{[\mu a}{}^b\partial_\nu w_{\rho]b}{}^a+\frac23w_{\mu a}{}^bw_{\nu b}{}^cw_{\rho c}{}^a\, ,
\ee
was defined in terms of the torsionful connection in flat index notation in \eqref{omega+-}.

\section{Bracket in MT scheme after reparameterization}\label{AppB}

As discussed in Subsection \ref{BracketRedef}, field redefinitions modify the brackets only through field dependent contributions in the zero order bracket. Starting from the symmetry algebra in the BdR scheme (\ref{brackets}) we observe that the only field dependent parameter is in
\begin{eqnarray}
\Lambda^{(0)}_{12 ab}\supset  
2 \, \beta_{1[a}{}^{c} \beta_{2\, b] c} = 
2 \, e_{\mu a} \, e_{\nu b}\, g_{\rho\sigma} \beta_{[1}^{\mu \rho} \; \beta_{2]}^{\sigma \nu} \;.
\end{eqnarray}

Hence, equation (\ref{BraFred}) implies that the only modification in the bracket, after the field redefinition that exchanges from the BdR to the MT scheme, is 
\begin{eqnarray}
\Lambda'_{12 ab} &=& \Lambda^{(old)}_{12 ab} -  \frac{a+b}{4}\left( H^{2}_{c[a}\beta^{[1}_{b]d} \beta^{2]cd} + H^{2cd} \beta^{[1}_{c[a} \beta^{2]}_{b]d}\right)\; .
\end{eqnarray}
Although it is not necessary a priori, in \eqref{transfmt} we performed an extra parameter redefinition, in order to put the $\beta$ transformations in a minimal form, where
\begin{eqnarray}
\Delta \zeta = \ell(\zeta,\psi) \to \left(\Delta \xi^{\mu}, \Delta \lambda_{a},\Delta\Lambda_{a b},\Delta \beta^{\mu\nu}\right)  = \left(0, -t_{a},-f_{ab},0\right) \;,
\end{eqnarray}
with $t_a$ and $f_{[ab]}$ defined in \eqref{fab} and \eqref{ta}.
$\Delta g_{12}$ is instead
\begin{eqnarray}
(\Delta g_{12})^{\mu\nu} &=& 0 \;,\\
(\Delta g_{12})^{\mu} 
&=& -\frac{a+b}{2} \beta_{1}^{\mu\nu} t_{2\nu}
=- \frac12 \beta_{1}^{\mu\nu} \beta_{2}^{ab}\omega_{\nu ab}=0\;,\\
(\Delta g_{12})_{\mu} 
&=&  2 \xi^{\nu}_{1} \partial_{[\mu}t_{2\nu]} \;,\\
(\Delta g_{12})_{ab} 
&=& - 2 \Lambda^{1}_{[a}{}^{c} f_{2b]c} 
- \xi_1^{\mu}\partial_{\mu} f_{2ab} \; \ .
\end{eqnarray}

Finally, we notice that $\hat \beta^{\mu\nu}_{12}=0,\; \hat\xi^{\mu}_{12}=0$, while 
\begin{eqnarray}
-2\hat\Lambda^{[12]}_{ab} &:=& \delta_{\xi_1,\lambda_1,\Lambda_1,\beta_1} \left(f_{2ab}\right) 
-  \delta_{\xi_2,\lambda_2,\Lambda_2,\beta_2} \left(f_{1ab}\right)\cr
&=& 2\; \xi^{[1\nu}\partial_{\nu}f_{2ab}
-\frac{a+b}{4} \beta^{[2}_{c[a}D_{b]}\Lambda^{1]}_{de}H^{cde} \cr
&&- (a+b) 
\; \beta_{[1}{}_{de} \beta_{2] c[a}  \left(\vphantom{\frac12}\right.
\omega_{b]f}{}^{d} \omega^{ef}{}_{c} 
+ \omega^{e}{}_{b]f} \omega_{c}{}^{fd}
- \frac12  H_{b]f}{}^{e} H^{df}{}_{c} 
\left.\vphantom{\frac12}\right)\cr
&&+  (a+b) 
\; \beta_{[1}{}_{c[a} \beta_{2]}{}^{cd}  \left(\vphantom{\frac12}\right.
R_{b]d} + 2 \nabla_{b]}\nabla_{d}\phi -\frac18 H^2_{b]d} -\frac12 \omega^2_{b]d}\left.\vphantom{\frac12}\right) \cr
&&+ (a+b)\; 
\beta_{[1}{}_{[a}{}^{c} \beta_{2]}{}_{b]}{}^{d}  \left(\vphantom{\frac12}\right.
R_{cd} + 2 \nabla_{c}\nabla_{d}\phi -\frac18 H^2_{cd} -\frac12 \omega^2_{cd}\left.\vphantom{\frac12}\right) \ ,
\end{eqnarray}
and
\begin{eqnarray}
-2\hat\lambda^{[12]}_{\nu} &=&  \delta_{\xi_1,\lambda_1,\Lambda_1,\beta_1} \left(t_{2\nu}\right)
- \delta_{\xi_2,\lambda_2,\Lambda_2,\beta_2} \left(t_{1\nu}\right) \cr
&=& 
4 \xi_{[1}^{\nu} \partial_{[\nu}{\hat t}_{2]\mu]}
+ 2(a+b)\; \left[
\beta_{[2}^{bc}\partial_{\mu}\Lambda_{1]bc}
+ \partial_{\mu}
\left( \xi_{[1}^{a}\beta_{2]}^{bc}\omega_{abc} \right)
+H_{\mu bc}\beta_{[1}^{bd} \beta_{2]}^{c}{}_{d} \right]\ .\label{hatl12}
\end{eqnarray}
The first term in (\ref{hatl12}) cancels with $(\Delta g_{12})_{\mu}$  and the remaining terms inside the bracket can be put in the form
\begin{eqnarray}
\left[\vphantom{\partial_{p}\beta^{mn}}\;\;\;\right] =
\frac12 \left(\beta_{[2}^{bc}\partial_{\mu}\Lambda_{1]bc}
-\partial_{\mu}\beta_{[2}^{bc}\; \Lambda_{1]bc}\right)
+ \partial_{\mu}
\left( \frac12 \beta_{[2}^{bc} \Lambda_{1]bc} + \xi_{[1}^{a}\beta_{2]}^{bc}\omega_{abc} 
+ b_{\nu\rho} \beta_{[1}^{\nu\sigma}\beta_{2]}^{\rho}{}_{\sigma} \right) \ ,
\end{eqnarray}
where the last factor is pure gauge and so can be ignored. Plugging these expressions with (\ref{newbracket}) one readily finds (\ref{bracketMT}).

\section{Leading order variation of  $L^{(1)}$ in MT scheme}\label{AppC}

 In this Appendix we present some details of the computations summarized in section \ref{MTsection} . The route  followed here is not unique, due to Bianchi identities and other less familiar relations, whose origin can be traced back to the constraints imposed after contractions with $\beta$. Following the structure of subsection \ref{MTsection}, we first deal with the even parity sector in \ref{l1+}, and then with the odd sector in \ref{l1-}.

\subsection{Variation of $L^{(1)}_+$}\label{l1+}

Here we present some details of the computations leading from equation \eqref{d0l+1}, namely 
\ba
\delta^{(0)}_\beta L_+^{(1)} &=&  \frac{a+b}{2} R^{abcd} \left[\nabla_{a}\left(\beta_{b}{}^{e} H_{e c d}\right) - \frac12 \nabla_{e}\beta_{ab} H^{e}{}_{cd}\right] \nn\\
&& - \frac{a+b}8 \;H_{f}{}^{ab} H^{fcd} \left[\nabla_{a}\left(\beta_{b}{}^{e} H_{e c d}\right) + \frac32 \nabla_{[e}\beta_{ab]} H^{e}{}_{cd}\right] \, ,\label{d0L1Bos}
\ea
to \eqref{VarAppendix}. 

A chain of integration by parts in the first term of the r.h.s. of (\ref{d0L1Bos}), followed by the Bianchi identity $\nabla_{[a}R_{bc]d}=0$ and a further integration by parts, leads to
\begin{eqnarray}
R^{abcd}\nabla_{a}\left(\beta_{b}{}^{e} H_{ecd}\right) &=&-2  R_{ab}\; \nabla_{c}\beta^{ae}\; H_{e}{}^{bc}
- 2 \beta^{e}{}_{a} R_{eb} {\cal D}_{c} H^{abc} 
+ 2\nabla_{a}\phi R^{abcd} \beta_{b}{}^{e} H_{ecd} \cr
&& + {\cal D}_{a}\left(R^{abcd} \beta_{b}{}^{e} H_{ecd} + 2 R_{bc} \beta^{be} H_{e}{}^{ca} \right)\; ,\label{AppAux1}
\end{eqnarray}
where, using $R_{abcd}\nabla^{d}\phi= 2\nabla_{[a}\nabla_{b]}\nabla_{c}\phi$ and integrating by parts, the third term in the r.h.s. can be rewritten as  
\begin{eqnarray}
2\nabla_{a}\phi R^{abcd} \beta_{b}{}^{e} H_{ecd} &=&
-4\nabla_{a}\nabla_{b}\phi \left(\nabla_{c}\beta^{a e} H_{e}{}^{bc}
 +{\cal D}_{c}H^{ebc}  \beta^{a}{}_{e}\right)
 +{\cal D}_{c}\left( 4 \nabla_{a}\nabla_{b}\phi \beta^{ae} H_{e}{}^{bc}   \right)\;.  \ \ \ \ \ \ \ \ \ \ \ \label{AppAux1B}
\end{eqnarray}
The second term in the r.h.s. of (\ref{d0L1Bos})  can be expressed alternatively as
\begin{eqnarray}
R^{abcd} \nabla_{e}\beta_{ab} H^{e}{}_{cd} &=&
-R^{abcd} \beta_{ab} {\cal D}_{e} H^{ecd} + {\cal D}_{e}\left( R^{abcd} \beta_{ab} H^{e}{}_{cd} \right)\;,
\end{eqnarray}
and the remaining terms  as
\begin{equation}\label{AppAux1C}
\begin{split}
\frac14 H_{f}{}^{ab} H^{fcd} &\left[\nabla_{a}\left(\beta_{b}{}^{e} H_{ecd}\right) +\frac32 \nabla_{[e} \beta_{ab]} H^{e}{}_{cd}\right] =\\ 
& =\frac12 H_{f}{}^{ab} H^{fcd}\left(  \nabla_{a}\beta_{b}{}^{e} H_{ecd} + \frac12\beta_{b}{}^{e} \nabla_{a}H_{ecd} +\frac14 \nabla_{e}\beta_{ab} H^{e}{}_{cd} \right) \;.
\end{split}
\end{equation}
Finally, the last two terms in \eqref{AppAux1C} can be recast in the form
\ba
 H_{f}{}^{ab} H^{fcd} \left(\beta_{b}{}^{e} \nabla_{a}H_{ecd} +\frac12 \nabla_{e}\beta_{ab} H^{e}{}_{cd} \right)
&=& - H^{abc} H_{bf}{}^{d} \left(2\beta^{ef}\nabla_{[e}H_{d]ca} - H_{aed} \nabla_{c}\beta^{ef} \right)\\
&=& - {\cal D}_{c}H^{abc} \beta^{ef} H_{bf}{}^{d} H_{aed} 
+ {\cal D}_{c}\left(\beta^{ef} H^{abc} H_{bf}{}^{d} H_{aed} \right)\, , \nonumber
\ea
where in the first equality we have used the relation 
\begin{eqnarray}
\frac12 H_{f}{}^{ab} H^{fcd} \nabla_{e}\beta_{ab} H^{e}{}_{cd} = 
- H^{abc} H_{bf}{}^{d} \left( \beta^{ef}\nabla_{e} H_{dca} - H_{aed} \nabla_{c}\beta^{ef}   \right)\;,\label{AppAux2}
\end{eqnarray}
while in the second line we have used the Bianchi identity $\nabla_{[a}H_{bcd]}=0$ before integrating by parts. The relation (\ref{AppAux2}) is not a Bianchi identity, but a consequence of the $\beta$ constraints, and it readily follows after explicit evaluation of the covariant derivatives.

Using (\ref{AppAux1})-(\ref{AppAux2}), we obtain the expression displayed in (\ref{VarAppendix}).
 
Let us pursue the computation further, to see how (\ref{var1}) is satisfied by (\ref{TransfPlus}) for $\gamma = \chi = 0$. Combining the above results with  (\ref{d1L0}) we obtain
\begin{eqnarray} 
\delta_{\beta}^{(1)} { L}^{(0)}+ \delta_\beta^{(0)} { L}^{(1)}_{+} = {\cal D}_{a} X^{a}\;,  \label{DX}
\end{eqnarray}
with
\begin{eqnarray}
X^{a} \!\!\! &=&\!\!\! - \frac {a+b} 2 \left[
  \nabla_{b}\phi \left( 3\nabla_{c}\beta^{b}{}_{d} H^{acd} 
- \nabla_{c}\beta^{a}{}_{d} H^{bcd} 
+ \beta_{c}{}^{d}\nabla_{d}H^{abc} \right)
-\beta_{be} H^{cd[e} R^{a]bcd} 
 \right. \label{AppAux3} \\
&-& \!\!\!\!
\left. 2\beta_{bc} R^{b}{}_{d} H^{acd}+\frac12 \nabla_{d}\beta_{c}{}^{a}\nabla_{b} H^{bcd} 
- \nabla_{c}\beta_{d}{}^{b} \nabla_{b} H^{acd} 
+\frac12\beta_{c}{}^{b} \nabla_{b}\nabla_{d} H^{acd}
-\frac14 \beta_{bc} H^{abd} H^{cef} H_{def} \right]\nn
 \end{eqnarray}
which can be cast in the form
\be
X^a = -\delta^{(0)}V^{a} + {\cal D}_{b}Z^{ab} \ , 
\label{AppAux3B}\ee
where 
 \begin{eqnarray}
 V^{a}=\frac{a+b}8 H^{acd}{\cal D}^{b} H_{bcd} \; , \;\;\;\; 
 Z^{ab}= - \frac {a+b} 2 \beta_{cd} \nabla^{c}H^{dab}\;.
 \end{eqnarray}
To put (\ref{AppAux3}) into the form (\ref{AppAux3B}) we used 
\begin{eqnarray}
\delta^{(0)}_\beta\nabla^{c}H_{cab}= - 4 R^{c}{}_{[a} \beta_{b]c} + \left(\beta_{de} H_{abc} + \beta_{e[a}H_{b]cd}\right) H^{cde}\;,
\end{eqnarray}
in addition to Riemann Bianchi identities. Noting that 
\begin{eqnarray}
{\cal D}_{a}{\cal D}_{b}Z^{ab}= \nabla_{a}\nabla_{b}Z^{ab} -2  \left(\nabla_{a}\nabla_{b}\phi -2 \nabla_{a}\phi\nabla_{b}\phi\right)Z^{ab}
- 2\left(\nabla_{a}\phi\nabla_{b}Z^{ab}+ \nabla_{b}\phi\nabla_{a}Z^{ab}\right) \ ,
\end{eqnarray}
vanishes for an antisymmetric object $Z^{ab}$, we get
\begin{eqnarray}
{\cal D}_{a} X^{a}=-{\cal D}_{a} \left(\delta^{(0)}_\beta V^a\right)=-\delta^{(0)}_\beta \left({\cal D}_{a}V^a\right)\;.
\end{eqnarray}
Finally, plugging this back into (\ref{DX}) we reproduce (\ref{var1}). This completes the discussion on the invariance of ${ L}^{(1)}_{+}$.

\subsection{Variation of $L^{(1)}_-$}\label{l1-}
Now we focus on the variation of ${L}^{(1)}_{-}$.   We start by analyzing the terms that are independent of the 3-form, namely
\begin{eqnarray}
-\frac14 \; \Omega^{abc} \delta^{(0)}_\beta H_{abc}  &=& 
-\frac18 \nabla_{a}\beta_{bc}w^{a}{}_{ef}\left(R^{bcef}
+ 2 \, w^{bf}{}_{d} w^{cd}{}_{e}\right)
-\frac14 \nabla_{b}\beta_{ca} w^{a}{}_{ef} R^{bcef}\;,\label{AppAux4}
\end{eqnarray}
where we have used 
\begin{eqnarray}
\Omega^{abc}=-\frac12 w^{[a}{}_{ef} R^{bc]ef} -\frac13 w^{ad}{}_{e} w^{be}{}_{f} w^{cf}{}_{d}\;.
\end{eqnarray}
The first term in the r.h.s. can be rewritten  as
\begin{eqnarray}
-\frac18 \nabla_{a}\beta_{bc}w^{a}{}_{ef}\left(R^{bcef}
+ 2 w^{bf}{}_{d} w^{cde}\right) &=&
-\frac14  \beta^{bc}w^{a}{}_{ef}\left( \nabla_{b}R_{ca}{}^{ef}
- 2 \nabla_{a}w_{b}{}^{fd} w_{cd}{}^{e}\right)\cr
&=& \frac14  \beta_{c}{}^{a} w_{abd} w^{d}{}_{ef}
\left(R^{bcef}+ 2 w^{beg}w^{cf}{}_{g}\right)\;,\label{nn}
\end{eqnarray}
where in the first equality we noted that 
\begin{eqnarray}
R^{bcef} + 2 w^{bf}{}_{d} w^{cde}
= 2D^{[b}w^{c]ef} + 2w^{[bc]d} w_{d}{}^{ef} \ ,
\end{eqnarray}
vanishes after contraction with $\beta_{bc}$, which allows  to move  the covariant derivative to the factors inside the parenthesis. Upon implementing Riemann Bianchi identities, the second equality in \eqref{nn} follows after applying the relations 
\begin{eqnarray}
\beta^{bc}\nabla_{a}w_{bde} &=& \beta^{bc} \left(R_{abde} + w_{ba}{}^{f} w_{fde} \right)\;,\;\;\; \textbf{\label{AppAux5A}}\\
\beta^{bc} \nabla_{b}R_{ca}{}^{ef} &=& - 2 \beta^{bc} w_{b}{}^{d[e} R_{cad}{}^{f]}\;. \label{AppAux5B}
\end{eqnarray}
The second term in the r.h.s. of (\ref{AppAux4}) can be cast as
\begin{eqnarray}
-\frac14 \nabla_{b}\beta_{ca} w^{a}{}_{ef} R^{bcef} = 
  \frac14  \beta_{ca} \nabla_{b}w^{a}{}_{ef} R^{bcef} 
-\frac14  \nabla_{b}\left( \beta_{ca} w^{a}{}_{ef} \right) R^{bcef}  \;.
\end{eqnarray}
Using (\ref{AppAux5A}), the first term in the r.h.s. of this expression can be rewritten  as
\begin{eqnarray}
\frac14  \beta_{ca} \nabla_{b}w^{a}{}_{ef} R^{bcef}= 
\frac14 \beta_{c}{}^{a} w_{ab}{}^{d} w_{def} R^{bcef}\;,
\end{eqnarray}
whereas the second one is 
\begin{eqnarray}
-\frac14 \nabla_{b}\left( \beta_{ca} w^{a}{}_{ef} \right) R^{bcef} 
 =-\frac14 {\cal D}_{b}\left[\beta_{ca} \left(w^{a}{}_{ef} R^{bcef} 
 - 2 w^{ab}{}_{d} (R^{cd} + 2 \nabla^{c}\nabla^{d}\phi)  \right) \right]\;.\label{AppAux5C}
\end{eqnarray}
Here we have followed similar steps to those performed in (\ref{AppAux1}) and (\ref{AppAux1B}), supplemented with the relation
\begin{eqnarray}
&&{\cal D}_{b}\left(\beta_{ca} w^{ab}{}_{d}\right)\left(R^{cd}+ 2\nabla^{c}\nabla^{d}\phi\right)
= \;\;\;\;\;\;\;\;\;\;\;\;\cr
&&\;\;\;\;\;\;\;\;\;\;\;\;\;\;\;\;\;\;\;\;\;\;
=\left[\beta_{c}{}^{a}\left(R^{a}{}_{d}+2\nabla^{a}\nabla_{d}\phi\right) +\beta_{ae} w^{ab}{}_{d} w^{e}{}_{bc}\right] \left(R^{cd}+ 2\nabla^{c}\nabla^{d}\phi\right)=0\; ,\;\;\;
\end{eqnarray}
which vanishes by symmetry arguments. Actually, using the relation
\begin{eqnarray}
{\cal D}_{b}\left[2\beta_{ca} w^{ab}{}_{d} \nabla^{c}\nabla^{d}\phi\right]=
{\cal D}_{b}\left[\nabla_{c}\left(\beta_{ae} w^{ac}{}_{d} w^{edb}\right)\right]\;,
\end{eqnarray}
it can be shown that (\ref{AppAux5C})  vanishes.

Regarding terms  quadratic  in the 3-form, we get
\begin{eqnarray}
-\frac14 H_{abc} \delta^{(0)} \Omega^{abc} &=& 
 \frac18 H^{abc} H^{e}{}_{af} \beta^{df} \; \left(R_{debc} -2 w_{be}{}^{g} w_{cdg}\right)\cr
&+& \frac18 H^{abc} H^{fde} \beta_{af} \;  w_{be}{}^{g} w_{cdg}\cr
&-&  \frac18 H^{abc} w _{a}{}^{de}\left[\nabla_{b}\left(\beta_{c}{}^{f} H_{fde}\right)
+ \nabla_{d}\left(\beta_{e}{}^{f} H_{fbc}\right)\right]\cr
&+&  \frac18 H^{abc} H_{ecf} w _{a}{}^{de}\left[\nabla_{b}\beta_{d}{}^{f} - \nabla_{d}\beta_{b}{}^{f} - \nabla^{f}\beta_{bd} \right]\;,
\label{AppAux6}
\end{eqnarray}
where we applied
\begin{eqnarray}
\beta_{a}{}^{f} H^{abc} H_{f}{}^{de} R_{debc}=0\, .
\end{eqnarray}
Using (\ref{AppAux5A}) and integrating by parts, the first line in \eqref{AppAux6} can be taken to the form 
\begin{eqnarray}
&& \frac18 H^{abc} H^{e}{}_{af} \beta^{df} \; \left(R_{debc} -2 w_{be}{}^{f} w_{cdf}\right)
= {\cal D}_{b}\left(-\frac18 H^{abc}H_{aef} \beta_{cd} w^{def}\right) \cr
&&\;\;\;\;\;\;\;\;\; - \frac18 {\cal D}_{b} H^{baf} \beta_{ca} H_{f}{}^{de} w_{cde} 
-\frac18 \nabla^{b} H^{adc} H_{fdc} \beta^{ef} w_{eba} 
+\frac12 H^{abc} H^{e}{}_{af} \beta^{df} w_{bd}{}^{g} w_{ceg}\;.\;\;\;\; \ \ \ \ \ \ \ \ \ 
\end{eqnarray}
The first term in the third line of (\ref{AppAux6}) is
\begin{eqnarray}
- \frac18 H^{abc} w _{a}{}^{de} \; \nabla_{b}\left(\beta_{c}{}^{f} H_{fde}\right) &=&
- \frac18 {\cal D}_{f}H^{fab} \beta_{ca} w_{b}{}^{de} H^{c}{}_{de} 
+  \frac18 \beta_{c}{}^{f} H^{abc} H_{fde} w _{a}{}^{dg} w_{b}{}^{e}{}_{g}\cr
&&
{\cal D}_{b}\left(- \frac18 H^{abc} w _{a}{}^{de} \beta_{c}{}^{f} H_{fde} \right)\;,
\end{eqnarray}
and the second one is 
\begin{eqnarray}
 -\frac18 H^{abc} w _{a}{}^{de} \nabla_{d}\left(\beta_{e}{}^{f} H_{fbc}\right)
 =
 \frac18 H^{abc} w _{ea}{}^{d} \nabla_{d}\left(\beta^{e f} H_{fbc}\right) 
- \frac14 H^{abc} w _{[ea]}{}^{d} \nabla_{d}\left(\beta^{e f} H_{fbc}\right)\;. \label{AppAux7}
\end{eqnarray}
Performing similar steps to those followed in (\ref{AppAux1})-(\ref{AppAux1B}), the first factor in (\ref{AppAux7}) can be cast  as
\begin{eqnarray}
\frac18 H^{abc} w _{ea}{}^{d} \nabla_{d}\left(\beta^{e f} H_{fbc}\right) 
&=& {\cal D}_{d}\left(\frac18 \beta^{ef} w_{ea}{}^{d} H^2{}^{a}{}_{f}\right) 
+\frac18 \beta^{ac} H^2{}^{b}{}_{c}\left(R_{ab} + \nabla_{a}\nabla_{b}\phi \right)\cr 
&&-
\frac18 \beta^{ef} H_{fbc} w_{ea}{}^{d} \nabla_{d}H^{abc} 
- \frac18  H^{2}{}^{a}{}_{f} \beta^{ef} w_{ed}{}^{g}w_{ga}{}^{d}\;.
\end{eqnarray}
Operating with the covariant derivative, followed by Bianchi identities of the 3-form  and integration by parts, the last component in (\ref{AppAux7}) can be worked out, leading to
\begin{eqnarray}
-\frac14 H^{abc} w _{[ea]}{}^{d} \nabla_{d}\left(\beta^{e f} H_{fbc}\right)&=& 
{\cal D}_{b}\left(-\frac12 H^{abc} w_{[ae]}{}^{d}\beta^{ef} H_{cdf} \right)
+ \frac12 {\cal D}_{f}H^{fab} H_{bcd}  w_{[ae]}{}^{c}\beta^{de} \nn\\
&& - \frac14 H^{abc} w_{[ea]d}\left( \beta^{g}{}_{f} w^{fde} H_{gbc}
+ 2\beta^{ef} w_{fb}{}^{g} H^d{}_{gc}  \right)  \\
&& - H^{abc} H_{cdf} \left( w_{[ea]}{}^{d} w_{gb}{}^{[e}\beta^{f]g} 
+  \frac14 \beta^{ef} (w_{ae}{}^{g}w_{bg}{}^{d} - w_{ea}{}^{g}w_{gb}{}^{d} )  \right).\nn
\end{eqnarray}
We notice that the last line in (\ref{AppAux6}) is simply
 \begin{eqnarray}
\frac18 H^{abc} H_{ecf} w _{a}{}^{de}\left[\nabla_{b}\beta_{d}{}^{f} - \nabla_{d}\beta_{b}{}^{f} - \nabla^{f}\beta_{bd} \right]&=&
-\frac14 \beta^{f}{}_{g} H^{abc} H_{ecf} w_{a}{}^{de} w^{g}{}_{db} \;.
\end{eqnarray}

Finally, plugging all these equations into (\ref{AppAux4}) and \eqref{AppAux6}, and after a proper rescaling with $a-b$, we obtain (\ref{d0LCS}). Note that we have  included for convenience a vanishing 
quartic factor 
\begin{eqnarray}
-\frac{1}{32}\beta_{ca}H^2_{b}{}^{c} H^2{}^{ab}=0\, .
\end{eqnarray}

We conclude this appendix by noticing that the total derivative from $\delta^{(1)}_{\beta}L^{(0)}$ in (\ref{d1L0}) with the choice $\kappa=\rho=0$ in (\ref{VarMenos}) exactly cancels the total derivative in (\ref{d0LCS}). This follows straightforwardly after using the relations 
\begin{eqnarray}
{\cal D}_{b}\left(\beta_{c}{}^{b} {\cal D}_{a}H^{2}{}^{ac}\right)=0\;,\;\;\;\;\;\;
{\cal D}_{b}\left(\beta_{c}{}^{b} \nabla_{a}w^{2}{}^{ac}\right)=0\;,\;\;\
\end{eqnarray}
which are particular cases of the more general identity valid for any $W^a$
\begin{eqnarray}
{\cal D}_{b}\left(\beta_{c}{}^{b} W^{c}\right)=0\;.
\end{eqnarray}

\end{appendix}

\end{document}